\documentclass[conference]{IEEEtran}
\IEEEoverridecommandlockouts

\usepackage{mathpazo} 
\usepackage[scaled=.80]{helvet}
\usepackage{courier}
\usepackage{times}
\usepackage{booktabs}
\usepackage{cite}
\usepackage{amsmath,amssymb,amsfonts}
\usepackage{algorithmic}
\usepackage{graphicx}
\usepackage{textcomp}
\usepackage{enumitem}
\usepackage{xcolor}
\usepackage{mathtools}
\usepackage{hyperref}

\def\BibTeX{{\rm B\kern-.05em{\sc i\kern-.025em b}\kern-.08em
    T\kern-.1667em\lower.7ex\hbox{E}\kern-.125emX}}
    
\newcommand{\nop}[1]{}

\newcommand*{\ie}{{\em i.e.}}    

\usepackage{tikz,amsmath, amssymb,bm,color}
\usetikzlibrary{shapes,arrows, arrows.meta}
\usetikzlibrary{arrows.meta,calc,positioning,patterns}

\usetikzlibrary{calc}

\tikzstyle{textnode} = [rectangle, inner sep=0pt,outer sep=0,execute at begin node={\strut}, font=\small]  
\tikzstyle{bnode} = [circle, draw, fill=black, minimum size=4mm, text=white, outer sep=1.5pt]  
\tikzstyle{enode} = [circle, draw, fill=gray!20, minimum size=4mm, inner sep=0.25pt, outer sep=1.5pt]  
\tikzstyle{inode} = [circle, thick, draw, minimum size=2mm, outer sep=1.5pt]  
\tikzstyle{nt} = [draw, inner xsep=1.5, fill=gray!5, minimum size=3mm, minimum size=1mm, outer sep=1.5pt]  
\tikzstyle{tnode} = [minimum size=5mm, font=\large]  
\tikzstyle{hnode} = [enode, very thick, draw=blue, outer sep=1.5pt]  
\tikzstyle{hidden} = [draw=none]
\tikzstyle{edge} = [->, thick, >=stealth', shorten >=1pt, auto,]  
\tikzstyle{iedge} = [edge, ultra thick, draw=blue]   
\tikzstyle{bedge} = [edge, draw=red]  
\tikzstyle{faded} = [opacity=0.60, text opacity=0.60]
\tikzstyle{Arrow} = [line width=0.5mm, draw=red!80, -{Triangle[length=1.5mm,width=1.5mm]}]
\tikzstyle{highlight} = [fill=gray!50, very thick]
\tikzstyle{treenode} = [enode, shrink]
\tikzstyle{tedge} = [->, >=latex, thin, draw=gray]
\tikzstyle{shrink} = [scale=0.5, transform shape]

\tikzstyle{textnode} = [rectangle, inner sep=0pt,outer sep=0,execute at begin node={\strut}, font=\small]  
\tikzstyle{bnode} = [circle, draw, fill=black, minimum size=4mm, text=white, outer sep=1.5pt]  
\tikzstyle{enode} = [circle, draw, fill=gray!20, minimum size=4mm, inner sep=0.25pt, outer sep=1.5pt]  
\tikzstyle{inode} = [circle, thick, draw, minimum size=2mm, outer sep=1.5pt]  
\tikzstyle{newnode} = [circle, thick, draw=blue, minimum size=2mm, outer sep=1.5pt, fill=white!80!blue]  
\tikzstyle{nt} = [draw, inner xsep=1.5, fill=gray!5, minimum size=3mm, minimum size=1mm, outer sep=1.5pt]  
\tikzstyle{tnode} = [minimum size=5mm, font=\large]  
\tikzstyle{hnode} = [enode, very thick, draw=blue, outer sep=1.5pt]  
\tikzstyle{hidden} = [draw=none]
\tikzstyle{edge} = [->, thick, >=stealth', shorten >=1pt, auto,]  
\tikzstyle{iedge} = [edge, ultra thick, draw=blue]   
\tikzstyle{bedge} = [edge, draw=red]  
\tikzstyle{uedge} = [thick, shorten >=1pt, auto,]  
\tikzstyle{faded} = [opacity=0.60, text opacity=0.60]
\tikzstyle{Arrow} = [line width=0.5mm, draw=red!80, -{Triangle[length=1.5mm,width=1.5mm]}]
\tikzstyle{highlight} = [fill=gray!50, very thick]
\tikzstyle{shrink} = [scale=0.5, transform shape]
\tikzstyle{treenode} = [enode, shrink]
\tikzstyle{tedge} = [->, >=latex, draw=gray, thin]
\tikzstyle{tedge2} = [-{Latex[length=1mm]}, draw=gray, thin]

\usepackage{pgfplots}
\usetikzlibrary{pgfplots.groupplots}
\selectcolormodel{cmyk}

\newenvironment{customlegend}[1][]{%
    \begingroup
    \csname pgfplots@init@cleared@structures\endcsname
    \pgfplotsset{#1}%
}{%
    \csname pgfplots@createlegend\endcsname
    \endgroup
}%

\def\addlegendimage{\csname pgfplots@addlegendimage\endcsname}

\pgfplotsset{
    jitter/.style={
        y filter/.code={\pgfmathparse{\pgfmathresult+rnd*#1}}
    },
    jitter/.default=0.05
}
\pgfplotsset{every tick label/.append style={font=\scriptsize}}
\pgfplotsset{compat=1.13}
\pgfdeclarelayer{background}
\pgfdeclarelayer{foreground}
\pgfsetlayers{background,main,foreground}   

\definecolor{TolDarkPurple}{HTML}{332288}
\definecolor{TolDarkBlue}{HTML}{6699CC}
\definecolor{TolLightBlue}{HTML}{88CCEE}
\definecolor{TolLightGreen}{HTML}{44AA99}
\definecolor{TolDarkGreen}{HTML}{117733}
\definecolor{TolDarkBrown}{HTML}{999933}
\definecolor{TolLightBrown}{HTML}{DDCC77}
\definecolor{TolDarkRed}{HTML}{661100}
\definecolor{TolLightRed}{HTML}{CC6677}
\definecolor{TolLightPink}{HTML}{AA4466}
\definecolor{TolDarkPink}{HTML}{882255}
\definecolor{TolLightPurple}{HTML}{AA4499}
\pgfplotscreateplotcyclelist{mbarplot cycle}{%
  {draw=TolDarkBlue,    fill=TolDarkBlue!70},
  {draw=TolLightBrown,  fill=TolLightBrown!70},
  {draw=TolLightGreen,  fill=TolLightGreen!70},
  {draw=TolDarkPink,    fill=TolDarkPink!70},
  {draw=TolDarkPurple,  fill=TolDarkPurple!70},
  {draw=TolDarkRed,     fill=TolDarkRed!70},
  {draw=TolDarkBrown,   fill=TolDarkBrown!70},
  {draw=TolLightRed,    fill=TolLightRed!70},
  {draw=TolLightPink,   fill=TolLightPink!70},
  {draw=TolLightPurple, fill=TolLightPurple!70},
  {draw=TolLightBlue,   fill=TolLightBlue!70},
  {draw=TolDarkGreen,   fill=TolDarkGreen!70},
}
\pgfplotscreateplotcyclelist{mlineplot cycle}{%
  {TolDarkBlue, mark=*, mark size=1.5pt},
  {TolLightBrown, mark=square*, mark size=1.3pt},
  {TolLightGreen, mark=triangle*, mark size=1.5pt},
  {TolDarkBrown, mark=diamond*, mark size=1.5pt},
}
\pgfplotsset{
  mlineplot/.style={
    mbaseplot,
    xmajorgrids=true,
    ymajorgrids=true,
    major grid style={dotted},
    axis x line=bottom,
    axis y line=left,
    legend style={
      cells={anchor=west},
      draw=none
    },
    cycle list name=mlineplot cycle,
  },
  mbarplot base/.style={
    mbaseplot,
    bar width=6pt,
  },
  mbarplot/.style={
    mbarplot base,
    ybar,
    xmajorgrids=false,
    ymajorgrids=true,
    area legend,
    legend image code/.code={%
      \draw[#1] (0cm,-0.1cm) rectangle (0.15cm,0.1cm);
    },
    cycle list name=mbarplot cycle,
  },
  horizontal mbarplot/.style={
    mbarplot base,
    xmajorgrids=true,
    ymajorgrids=false,
    xbar stacked,
    area legend,
    legend image code/.code={%
      \draw[#1] (0cm,-0.1cm) rectangle (0.15cm,0.1cm);
    },
    cycle list name=mbarplot cycle,
  },
  mbaseplot/.style={
    legend style={
      draw=none,
      fill=none,
      cells={anchor=west},
    },
    x tick label style={
      font=\footnotesize
    },
    y tick label style={
      font=\footnotesize
    },
    legend style={
      font=\footnotesize
    },
    major grid style={
      dotted,
    },
  },
  disable thousands separator/.style={
    /pgf/number format/.cd,
      1000 sep={}
  },
}

\begin{document}

\title{Towards Interpretable Graph Modeling \\ with Vertex Replacement Grammars
}

\author{\IEEEauthorblockN{Justus Hibshman \qquad Satyaki Sikdar \qquad Tim Weninger} 
\IEEEauthorblockA{Department of Computer Science \& Engineering \\ 
University of Notre Dame \\ 
Notre Dame, IN, USA\\
\texttt{\{jhibshma,ssikdar,tweninge\}@nd.edu}} 
} 

\maketitle

\begin{abstract}
An enormous amount of real-world data exists in the form of graphs. Oftentimes, interesting patterns that describe the complex dynamics of these graphs are captured in the form of frequently reoccurring substructures. Recent work at the intersection of formal language theory and graph theory has explored the use of graph grammars for graph modeling and pattern mining. However, existing formulations do not extract meaningful and easily interpretable patterns from the data. The present work addresses this limitation by extracting a special type of vertex replacement grammar, which we call a KT grammar, according to the Minimum Description Length (MDL) heuristic. In experiments on synthetic and real-world datasets, we show that KT-grammars can be efficiently extracted from a graph and that these grammars encode meaningful patterns that represent the dynamics of the real-world system.
\end{abstract}


\section{Introduction}\label{AA}
A common task in big data is to seek and find patterns hidden in enormous amounts of data. When the data takes the form of the graph, this goal is expressed as finding meaningful graphical substructures and other patterns that are hidden in the graph. Because of the prevalence of graph data and the importance of this task, dozens of graph models have been developed towards this goal~\cite{ahmed2015efficient,seshadhri2012community,aguinaga2016growing,aguinaga2018learning}. Typically, these graph models make some assumptions about the shape or structure of the graph and encode the graph in interesting ways. 

Some of the most widely used graph modeling techniques search for occurrences of specific structures, such as edges, triangles, various 4-node graphlets, and so on. Other techniques measure specific graph properties such as node centrality, degree, or measures of network robustness. What almost all of these methods have in common is the fact that they typically learn structures that are specified in advance~\cite{koutra2014vog}.

We currently lack modeling tools that allow the graph itself to dictate which graph patterns are essential and then report these newfound properties in a meaningful and human-readable format.

A few approaches are closer to this ideal than most. Notable works such as gSpan~\cite{yan2002gspan}, CloseGraph~\cite{yan2003closegraph}, and SUBDUE~\cite{holder1994substucture} search for arbitrary substructures in a graph that can be used to create a lossy compression of the graph. However, these existing tools do little to show how those structures connect to each other and the rest of the graph, and some of them have trouble scaling to large or even mid-sized graphs. Progress in graph entropy uses an information-theoretic approach to identify graph structures and is a promising direction but does not produce an interpretable model~\cite{gudkov2016generalized}.

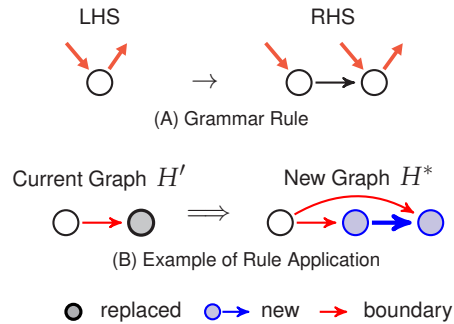
\begin{figure}
    \centering
    \begin{tikzpicture}
\begin{scope}[shift={(0,0.3)}]

\node [inode] (v2) at (-2.65,2.75) {};
\node [hidden] (v1) at (-3.2,3.4) {};
\node [hidden] (v3) at (-2.2,3.4) {};
\draw [Arrow] (v1) -- (v2);
\draw [Arrow] (v2) -- (v3);

\node at (-1.25,2.75) {$\rightarrow$};

\node [inode] (v4) at (0,2.75) {};
\node [inode] (v5) at (1,2.75) {};
\node [hidden] (v3) at (1.47,3.4) {};
\node [hidden] (v1) at (0.45,3.4) {};

\node [hidden] (v6) at (-0.55,3.4) {};

\draw [edge] (v4) edge (v5);
\draw [Arrow] (v5) -- (v3);
\draw [Arrow] (v1) -- (v5);
\draw [Arrow] (v6) -- (v4);
\node at (-2.65,3.625) {\textsf{LHS}};
\node at (0.45,3.625) {\textsf{RHS}};

\node [textnode] at (-0.9,2.25) {\textsf{(A) Grammar Rule}};
\end{scope}

\begin{scope}[shift={(-7.65,-0.2142)}]
\node [inode, ] (v1) at (4.55,1.4) {};
\node [inode, highlight] (v2) at (5.55,1.4) {};
\draw [bedge] (v1) -- (v2);

\node [] at (6.45,1.5) {$\Longrightarrow$};
\node [inode, ] (v1) at (7.4,1.4) {};
\node [newnode] (v2) at (8.4,1.4) {};
\node [newnode] (v3) at (9.4,1.4) {};
\draw [iedge] (v2) -- (v3);
\draw [bedge] (v1) -- (v2);
\draw [bedge] (v1) edge[bend left, draw=red] (v3);

\node [textnode] at (6.95,0.9) {\textsf{(B) Example of Rule Application}};
\node at (5,2) {\textsf{Current Graph} $H^\prime$};
\node at (8.45,2) {\textsf{New Graph} $H^*$};
\end{scope}

\begin{scope}[shift={(0.7,-5.5)}]
\node [hidden] (v7) at (-0.55,5.5) {};
\node [hidden] (v8) at (0.1,5.5) {};
\draw [bedge] (v7) edge[draw=red] (v8);

\node [hidden] at (0.75,5.5) {\textsf{boundary}};
\node[circle, scale=0.6, very thick, draw=black, fill=gray!75] at (-3.7,5.5) {};
\node [hidden] (v8) at (-1.2,5.5) {};

\node [hidden] at (-0.95,5.5) {\textsf{new}};

\node[newnode, scale=0.6] (n) at (-1.85,5.5) {};
\draw [iedge, thick] (n) edge[draw=blue] (v8);
\node [hidden] at (-2.8,5.5) {\textsf{replaced}};

\end{scope}

\end{tikzpicture}
    \caption{(A) Example KT-grammar production rule with a left-hand side (LHS) and a right-hand side (RHS). The LHS is a single node with zero or more incoming and outgoing boundary edges (drawn in red). The RHS is a subgraph fragment, where each vertex has zero or more incoming and outgoing boundary edges. (B) During generation, a vertex from the graph is replaced by the RHS; incoming and outgoing boundary edges from the LHS are rewired to \textit{all} of the incoming and outgoing boundary edges of the RHS respectively.}
    \label{fig:rule}
\end{figure}

Renewed interest in graph grammars provides a promising route towards the goal of building a non-parametric, interpretable graph model. Previous work has investigated the relationship between graph mining and formal language theory by extracting Vertex Replacement Grammars (VRGs)~\cite{sikdar2019modeling} and (Hyper)edge Replacement Grammars (HRGs)~\cite{aguinaga2016growing,reddy2019edge}. Unfortunately, the composition of grammar rules in HRGs, and some VRGs are known to produce clunky patterns that are difficult to interpret.

The present work uses the graph grammar introduced by Kemp and Tennenbaum (KT), which originally included a Bayesian graph model that could learn natural relationships between items in tiny datasets~\cite{kemp2008discovery}. Generally speaking, KT-grammars, as we call them, are based on prior work in vertex replacement grammars, which contain graphical rewriting rules that can match and replace graph fragments similar to how a context-free string grammar rewrites characters in a string~\cite{ehrig1999handbook}. These graph fragments represent a succinct description of the building blocks of the network, and the rewiring rules of the grammar describe the instructions about how the graph is pieced together.

KT-grammars, which are a specific type of VRGs, are used to model graph structures and can even generate new graphs. A KT-grammar rule replaces a single \textit{vertex} with a subgraph fragment as shown in Fig~\ref{fig:rule}. 
KT-grammars are easy to use and easy to interpret, but their current use requires human modelers to craft these grammars by hand, which is time consuming and introduces human biases into the model. The rule inference system developed by Kemp and Tennenbaum has shown some promise in determining which rules best match data, but this system is limited to datasets of only a few dozen items~\cite{kemp2008discovery}. We desire an automatic, scalable, and interpretable rule extraction algorithm that compactly models the structures found in the graph.

To that end, the present work describes BUGGE: a \textbf{B}ottom-\textbf{u}p \textbf{G}raph \textbf{G}rammar \textbf{E}xtractor (pronounced: ``buggie''), which extracts interpretable KT-grammars from large real-world graphs. We show that the KT-grammar and the BUGGE extractor can correctly capture the known generative process of synthetic graphs. Based on their success in synthetic graphs, we employ BUGGE to find hidden structures in real-world graphs and report the findings.

\section{Preliminaries}

Before we describe BUGGE in detail, we first give some important background information. The BUGGE algorithm can take, as input, any graph $H = (V, E)$, which can be labeled, weighted, multi-edged, or directed. 
However, for simplicity, our implementation and the examples presented in this paper focus on simple, directed graphs with no edge weights or labels. Note that we use ``vertex'' and ``node'' interchangeably.

\subsection{Vertex Replacement Grammars}

A vertex replacement grammar is a context-free graph grammar consisting of a set of ``production rules.'' These production rules (or simply ``rules'') prescribe a way to replace a single vertex in the graph with a subgraph fragment. When a vertex replacement occurs, the orphaned edges adjacent to the deleted vertex needs to be rewired to the new subgraph fragment in some way. Various edge rewiring schemes have been developed, each with their advantages and disadvantages. 

\vspace{.2cm}
\noindent\textbf{The KT-grammar.}
The vertex replacement grammar introduced by Kemp and Tennenbaum is a natural formalism for our purposes~\cite{kemp2008discovery}. This formalism, which we call a KT-grammar, is succinct and easy to interpret, but it is also rigid and sometimes requires algorithmic tradeoffs.

Formally, a KT-Grammar $G$ is defined as a set of rules $R \in G$. Let $R = (F, i, o, f)$ such that $F = (V_R, E_R)$ is a directed graph fragment with vertices $v\in V$ and edges $e\in E$, $i: V_R \mapsto \{0, 1\}$ and $o: V_R \mapsto \{0, 1\}$ are indicator functions that state whether a vertex has incoming ($i$) or outgoing ($o$) boundary edges. $f \in \mathbb{Z}^+$ is the rule's ``frequency,'' a count of how many times that rule occurs.

Let $H = (V, E)$ be a directed graph upon which $G$ is applied. Vertex replacement is defined as a transformation of $H$ from a previous state $H^\prime$ to a new state $H^\ast$ via $R \in G$. Let $v \in V$ be the vertex in $H$ replaced by grammar rule $R = (F = (V_R, E_R), i, o, f)$. Then $H^\ast = (V^\ast, E^\ast)$, where the new vertices are:
\begin{align}
    V^\ast =&\ (V^\prime\, \setminus\, \{v\})\  \bigcup\ V_R
\shortintertext{and the new edges are:}
    E^\ast =&\ \{(s, t)\ |\ \  s,\,t \ne v \land ((s, t) \in E^\prime)\ \lor \nonumber \\ 
    & \qquad \qquad (t \in V_R \land i(t) \land (s, v) \in E^\prime)\ \lor \nonumber \\ 
    & \qquad \qquad (s \in V_R \land o(s) \land (v, t) \in E^\prime)\}
\end{align}
Simply put, whenever a rule replaces a node $x$, every node in $R$ either gets all of $x$'s boundary edges or none of them.

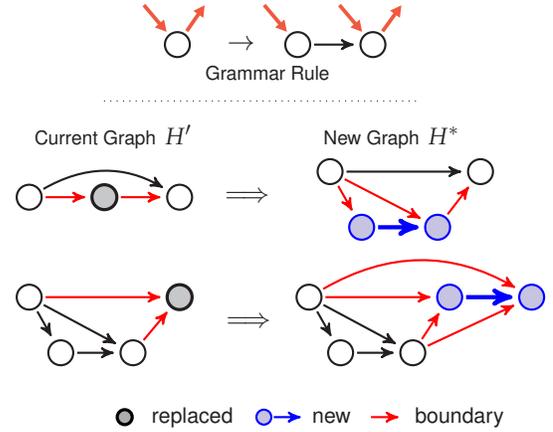
\begin{figure}
    \centering
    \begin{tikzpicture}
\node [textnode] at (-3.5,2.25) {\textsf{Current Graph} $H^\prime$};

\node [textnode] at (0.2,2.25) {\textsf{New Graph} $H^\ast$};



\begin{scope}[shift={(-11.82,-0.53)}]
\node [inode, ] (v1) at (7.2,2) {};
\node [inode, highlight] (v2) at (8.2,2) {};
\node [inode, ] (v3) at (9.2,2) {};

\draw [bedge] (v1) -- (v2);
\draw [bedge] (v2) -- (v3);
\draw [edge] (v1) edge[bend left, ] (v3);

\node [] at (10.1,2) {$\Longrightarrow$};
\node [inode, ] (v1) at (11.2,2.34) {};
\node [newnode] (v2) at (11.62,1.6) {};
\node [newnode] (v3) at (12.63,1.6) {};
\node [inode, ] (v4) at (13.2,2.34) {};

\draw [edge] (v1) -- (v4);
\draw [iedge] (v2) -- (v3);
\draw [bedge] (v1) -- (v2);
\draw [bedge] (v1) -- (v3);
\draw [bedge] (v3) -- (v4);
\end{scope}

\begin{scope}[shift={(-16.22,-2.2)}]
\node [inode, ] (v1) at (11.6,2.34) {};
\node [inode,] (v2) at (12.02,1.6) {};
\node [inode,] (v3) at (12.98,1.6) {};
\node [inode, highlight] (v4) at (13.6,2.34) {};

\draw [bedge] (v1) -- (v4);
\draw [edge] (v2) -- (v3);
\draw [edge] (v1) -- (v2);
\draw [edge] (v1) -- (v3);
\draw [bedge] (v3) -- (v4);

\node [] at (14.52,2) {$\Longrightarrow$};

\node [inode, ] (v1) at (15.32,2.34) {};
\node [inode, ] (v2) at (15.74,1.6) {};
\node [inode, ] (v3) at (16.7,1.6) {};
\node [newnode] (v4) at (17.19,2.34) {};
\node [newnode] (v5) at (18.28,2.34) {};

\draw [edge] (v2) -- (v3);
\draw [edge] (v1) -- (v2);
\draw [edge] (v1) -- (v3);
\draw [iedge] (v4) -- (v5);
\draw [bedge] (v3) -- (v4);
\draw [bedge] (v1) -- (v4);
\draw [bedge] (v3) -- (v5);
\draw [bedge] (v1) edge[bend left, draw=red] (v5);
\end{scope}

\begin{scope}[shift={(-0.4,0.75)}]

\node [inode] (v2) at (-2.25,2.75) {};
\node [hidden] (v1) at (-2.8,3.4) {};
\node [hidden] (v3) at (-1.8,3.4) {};
\draw [Arrow] (v1) -- (v2);
\draw [Arrow] (v2) -- (v3);

\node at (-1.4,2.75) {$\rightarrow$};

\node [inode] (v4) at (-0.65,2.75) {};
\node [inode] (v5) at (0.35,2.75) {};
\node [hidden] (v3) at (0.82,3.4) {};
\node [hidden] (v1) at (-0.2,3.4) {};

\node [hidden] (v6) at (-1.2,3.4) {};

\draw [edge] (v4) edge (v5);
\draw [Arrow] (v5) -- (v3);
\draw [Arrow] (v1) -- (v5);
\draw [Arrow] (v6) -- (v4);
\end{scope}

\node [textnode] at (-1.45,3.1) {\textsf{Grammar Rule}};

\node [hidden] (h1) at (-3.75,2.75) {};
\node [hidden] (h2) at (0.7,2.75) {};
\draw [dotted] (h1) -- (h2);

\begin{scope}[shift={(0.35,-6.9563)}]
\node [hidden] (v7) at (-0.55,5.5) {};
\node [hidden] (v8) at (0.1,5.5) {};
\draw [bedge] (v7) edge[draw=red] (v8);

\node [hidden] at (0.75,5.5) {\textsf{boundary}};
\node[circle, scale=0.6, very thick, draw=black, fill=gray!75] at (-3.7,5.5) {};
\node [hidden] (v8) at (-1.2,5.5) {};

\node [hidden] at (-0.95,5.5) {\textsf{new}};

\node[newnode, scale=0.6] (n) at (-1.85,5.5) {};
\draw [iedge, thick] (n) edge[draw=blue] (v8);
\node [hidden] at (-2.8,5.5) {\textsf{replaced}};

\end{scope}

\end{tikzpicture}
    \caption{A grammar rule repeatedly applied to grow a graph. At the top is the grammar rule; boundary edges, illustrated in red, indicate how the new RHS is rewired to the overall graph $H$. A red edge in a rule stands for connections to \textit{all} a vertex's neighbors (0 or more). The next two rows show this rule applied to grey vertices in a $H^\prime$. The expanded graph $H^\ast$ is illustrated on the right where the rule's RHS is highlighted in blue.}
    \label{fig:single-rule-example}
\end{figure}

The example in Fig.~\ref{fig:single-rule-example} shows two additional applications of the rule from Fig.~\ref{fig:rule}A. This single KT-grammar rule can be represented formally as $(F = (\{x, y\},\, \{(x, y)\}),\, i,\, o,\, f)$ where $i(x) = 1,\, o(x) = 0,\, i(y) = 1$, $o(y) = 1$, and $f$ is some positive integer.

Note that if a grammar rule has $i(v) = 0$ for all $v \in V$, then it cannot be used to replace a node with incoming edges. Likewise for outgoing boundary edges. Thus, the grammar rule's $i$ and $o$ functions implicitly define the left-hand side (LHS) of the rule. KT Grammar rules can have any number of nodes on the right-hand side (RHS).

\subsection{Minimum Description Length Principle}

The Minimum Description Length (MDL) principle asserts that the best representation of some data is the representation that uses the fewest bits. While this may be a questionable assertion philosophically, practically it is often a useful principle for big data mining and modeling. For example, gap-encodings can represent sparse matrices much more efficiently than a direct ``matrix'' encoding because gap encodings better ``match'' the data\cite{elias1975universal}.

The MDL principle may also be used in the following way: Given some data $D$, a set of models $\mathcal{M}$, and a particular encoding scheme, the best model to encode $D$ is the model $M \in \mathcal{M}$ that minimizes the combined cost of encoding $D$ given $M$ and the cost of encoding $M$ ($E(D | M) + E(M)$). 

In the present work, our data will be a graph, and our set of models will be a set of different grammar rules which our algorithm discovers. We will repeatedly, greedily select a grammar rule to compress the graph according to the MDL principle.

\section{Extracting KT-Grammars}\label{sec:grammar-extract}

In this section, we describe BUGGE, and show that it can extract a succinct, meaningful KT-grammar that faithfully represents the graphical structures and properties of large graph data. We introduce BUGGE formally and then describe how it works using a running example. 


\subsection{BUGGE: the Bottom-Up Graph Grammar Extractor}\label{subsec:BUGGE}

Let $H = (V, E)$ be a directed graph with $V^\prime \subseteq V$, and let $H(V^\prime)$ denote the subgraph in $H$ induced by $V^\prime$. We also introduce two size parameters $k_{\text{min}}$ and $k_{\text{max}}$ that bound the size of rule fragments. Let our grammar $G$ start as an empty set.

At a high-level BUGGE extracts a vertex replacement grammar in the following way:

\begin{enumerate}[label=Step~\arabic*,leftmargin=*,align=left]
\item Find all connected sets of nodes meeting the size constraints $k_\text{min} \leq $ size $\leq k_\text{max}$.

\item For each connected set of nodes $V^\prime$, find the rules $G'$ that could be used in reverse to contract $V^\prime$ into a single node. If no rules match exactly, find the rules which most closely match.
\item\label{alg-simple:SelectRule} Pick the single grammar rule $R$ which is predicted to compress $H$ the most.
\item\label{alg-simple:CollapseRule} Extract an occurrence of $R\in G^\prime$ from $H$ by applying it in reverse. If $R$ does not exactly match the nodes it collapses together, adjust the graph to make it fit (\ie, add or delete edges). If $R$ is not in our grammar $G$, add it to $G$. Increment the frequency of $R$ in $G$.
\item Update the sets of connected nodes and the associations of vertex sets to rules according to the new graph. Repeat \ref{alg-simple:SelectRule} and \ref{alg-simple:CollapseRule} if there are still rules which can be extracted.
\end{enumerate}

\noindent This principled approach extracts a vertex replacement grammar and can be applied to any kind of graph or grammar formalism. However, our goal of extracting a small, easily interpretable model is best satisfied by the KT-grammar formalism given earlier. So in the remainder of this section, we provide further details on how to extract a KT-grammar specifically.

\vspace{.2cm}
\noindent\textbf{Step 1: Enumerating Occurrences of Rules.}
When BUGGE first starts, it must enumerate rule occurrences for any connected set meeting the size constraints. Later, BUGGE only needs to perform updates to sets which might have been affected by the latest rule extraction (i.e., sets connected to nodes used in the latest extraction).


Enumerating all connected sets of nodes up to a fixed size ($k_\text{max}$ in our case) corresponds exactly to enumerating all connected, induced subgraphs up to some fixed size, which is solved using a technique called Reverse Search \cite{avis1996reverse}.

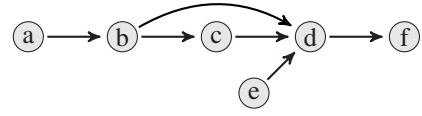
\begin{figure}
    \centering
    \begin{tikzpicture}
\begin{scope}
\node [enode] (a) at (-4.75,3.5) {a};
\node [enode] (b) at (-3.5,3.5) {b};
\node [enode] (c) at (-2.25,3.5) {c};
\node [enode] (d) at (-1,3.5) {d};
\node [enode] (e) at (-1.75,2.75) {e};
\node [enode] (f) at (0.25,3.5) {f};

\draw [edge] (a) -- (b);
\draw [edge] (b) -- (c);
\draw [edge] (b) edge[bend left, draw=black] (d);
\draw [edge] (c) -- (d);
\draw [edge] (d) -- (f);
\draw [edge] (e) -- (d);
\end{scope}
\end{tikzpicture}
    \caption{Example directed graph. This graph will be used as a running example to show how BUGGE extracts KT-grammar rules.}
    \label{fig:running_example}
\end{figure}

To show this process we introduce a running example using the graph shown in Fig.~\ref{fig:running_example}. With $k_\text{min}$ = $k_\text{max}$ = $2$, there exists one connected subgraph per edge, 6 in total: \{a,b\}, \{b,c\}, \{b,d\}, \{c,d\}, \{d,f\}, and \{e,d\}. With $k_\text{max} = 3$, we find 8 additional subgraphs: \{a,b,c\}, \{a,b,d\}, \{b,c,d\}, \{b,d,e\}, \{b,d,f\}, \{c,d,e\}, \{c,d,f\}, and \{e,d,f\}. The total number of sets for a graph tends to be exponential in $k_\text{max}$. 

Fortunately, the KT-grammar permits heuristics that allow the connected subgraph enumeration to be stopped early in many cases. If the enumerator just evaluated some set of nodes $X$ of size $|X| < k_\text{max}$ and we can infer that it is unlikely for any KT-Grammar rule including the nodes in $X$ to be a ``good'' rule (more on this in Step 3), then the search can ignore more massive sets of nodes containing $X$. We use this ``Enumeration Heuristic'' in our experiments to speed up computation while retaining results of similar quality.

\vspace{.2cm}
\noindent\textbf{Step 2: Enumerating Rules that Apply to Subgraphs.} Given the collection of connected sets, BUGGE finds the grammar rule(s) which best match each connected set.

For any given connected set, we consider every possible arrangement of boundary edges. In other words, for a given connected set $V^\prime$, we consider $R = (F, i, o, f)$ for every possible $i$ and $o$, where $F$ is the induced subgraph $H(V^\prime)$. For each rule (each possible $i, o$ pair), we consider the minimum number of edge additions or deletions to $H$ necessary to make $V^\prime$ correspond to an occurrence of $R$. This number of modifications is the ``cost'' of a rule occurrence. For a given $V^\prime$, we only store the rules with the lowest cost. Note that there are $2^{|V^\prime|}$ possible $i$ functions (and the same for $o$). Thus, this process is exponential in $k_\text{max}$.

\begin{figure}
    \centering
    \begin{tikzpicture}

\begin{scope}[shift={(-0.25,0.58)}, scale=0.8]
\draw[ thick, draw=orange, fill=orange!60]  (-1.6248,3.5) ellipse (1 and 0.42);
\node [enode, faded] (a) at (-4.75,3.5) {a};
\node [enode] (b) at (-3.5,3.5) {b};
\node [enode] (c) at (-2.25,3.5) {c};
\node [enode] (d) at (-1,3.5) {d};
\node [enode] (e) at (-1.75,2.65) {e};
\node [enode] (f) at (0.25,3.5) {f};

\draw [edge, faded, opacity=0.2] (a) -- (b);
\draw [bedge] (b) -- (c);
\draw [bedge] (b) edge[bend left, draw=red] (d);
\draw [edge] (c) -- (d);
\draw [bedge, dotted] (e) -- (d);
\draw [bedge] (d) -- (f);
\end{scope}

\node[draw=none] at (0.875,3.2) {$\longrightarrow$};
\node[draw=none] at (2.675,4.2) {\textsf{Rule 2}};

\begin{scope}[shift={(3.825,0.745)}, scale=0.85]
\node [inode] (v2) at (-2.25,2.75) {};
\node [hidden] (v1) at (-2.8,3.4) {};
\node [hidden] (v3) at (-1.8,3.4) {};
\node [textnode] at (-1.15,3.55) {\{b\}};
\node [textnode] at (0.8,3.575) {\{f\}};

\draw [Arrow] (v1) -- (v2);
\draw [Arrow] (v2) -- (v3);

\node at (-1.4,2.75) {$\rightarrow$};

\node [inode] (v4) at (-0.65,2.75) {};
\node [inode] (v5) at (0.35,2.75) {};
\node [hidden] (v3) at (0.82,3.4) {};
\node [hidden] (v1) at (-0.2,3.4) {};
\node [hidden] (v6) at (-1.2,3.4) {};

\node [textnode] at (-2.8,3.55) {\{b\}};
\node [textnode] at (-0.25,3.575) {\{b\}};
\node [textnode] at (-1.85,3.575) {\{f\}};
\draw [edge] (v4) edge (v5);
\draw [Arrow] (v5) -- (v3);
\draw [Arrow] (v1) -- (v5);
\draw [Arrow] (v6) -- (v4);
\end{scope}

\begin{scope}[shift={(-0.25,2.75)}, scale=0.8]
\draw[ thick, draw=orange, fill=orange!60]  (-1.6248,3.5) ellipse (1 and 0.42);
\node [enode, faded] (a) at (-4.75,3.5) {a};
\node [enode] (b) at (-3.5,3.5) {b};
\node [enode] (c) at (-2.25,3.5) {c};
\node [enode] (d) at (-1,3.5) {d};
\node [enode] (e) at (-1.75,2.65) {e};
\node [enode] (f) at (0.25,3.5) {f};

\draw [edge, faded, opacity=0.2] (a) -- (b);
\draw [bedge, dotted] (b) -- (c);
\draw [bedge] (b) edge[bend left, draw=red] (d);
\draw [edge] (c) -- (d);
\draw [bedge] (e) -- (d);
\draw [bedge] (d) -- (f);

\end{scope}
\node[draw=none] at (0.875,5.38) {$\longrightarrow$};
\node[draw=none] at (2.675,6.35) {\textsf{Rule 1}};

\begin{scope}[shift={(3.825,2.895)}, scale=0.85]
\node [inode] (v2) at (-2.25,2.75) {};
\node [hidden] (v1) at (-2.78,3.4) {};
\node [hidden] (v3) at (-1.8,3.4) {};
\node [textnode] at (-0.25,3.575) {\{b,e\}};
\node [textnode] at (0.775,3.55) {\{f\}};

\draw [Arrow] (v1) -- (v2);
\draw [Arrow] (v2) -- (v3);

\node at (-1.4,2.75) {$\rightarrow$};

\node [inode] (v4) at (-0.65,2.75) {};
\node [inode] (v5) at (0.35,2.75) {};
\node [hidden] (v3) at (0.8,3.4) {};
\node [hidden] (v1) at (-0.2,3.4) {};
\node [textnode] at (-2.8,3.55) {\{b,e\}};
\node [textnode] at (-1.825,3.55) {\{f\}};

\draw [edge] (v4) edge (v5);
\draw [Arrow] (v5) -- (v3);
\draw [Arrow] (v1) -- (v5);
\end{scope}

\node [textnode] at (-4.1,6.3) {\large\textsf{A}};
\node [textnode] at (-4.1,4) {\large\textsf{B}};
\end{tikzpicture}
    \caption{Two options for extracting a KT-grammar rule from vertices \{c, d\} highlighted in orange. Red dotted edges indicate edge deletions necessary to make the graph match the rule on the right; edges deleted (or added) incur a cost to the model. Solid red edges are the preserved boundary edges. Labeled edges in the rules are for illustrative purposes only.}
    \label{fig:rule_extract}
\end{figure}
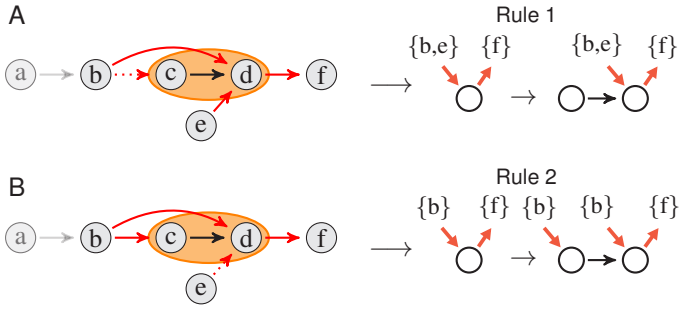

Figure~\ref{fig:rule_extract} contains an example of finding matching rules for a connected subgraph induced by c and d. It illustrates two different rules that the subgraph could be edited to. Figure~\ref{fig:rule_extract}A shows that the boundary edge (b$\rightarrow$c) does not match Rule 1, resulting in a cost of 1 for Rule 1. Figure~\ref{fig:rule_extract}B shows that the boundary edge (e$\rightarrow$d) does not match Rule 2 resulting in a cost of 1 for Rule 2.

To make a decision in Step 3, BUGGE needs to aggregate information on all the occurrences of an individual rule. To do this, it assigns an id number to every rule. We begin with an empty sequence of discovered rules, a ``rule library'', $L = \langle \rangle$. Every time an occurrence of a rule $R$ is found in $H$, we check to see if $R$ is isomorphic to a rule already in $L$. If not, we append it to $L$ and give $R$ a new id number. Otherwise, we give $R$ the id of its match in $L$. To make this process more efficient, we maintain a count of how many times each rule has been discovered and adjust the order of rules in $L$ to be in the order of discovery count, thereby increasing the likelihood that a newly discovered rule will match one of the first few rules in $L$.

\nop{
\begin{figure}
    \centering
    \begin{tikzpicture}
\begin{scope}[shift={(-0.4,0)}]
\node [inode] (v2) at (-2.25,2.75) {};
\node [hidden] (v3) at (-1.8,3.4) {};
\draw [Arrow] (v2) -- (v3);

\node at (-1.4,2.75) {$\rightarrow$};


\node [inode] (v4) at (-0.65,2.75) {};
\node [inode] (v5) at (0.35,2.75) {};
\node [hidden] (v3) at (0.82,3.4) {};
\node [textnode] at (-2.96,2.74) {$1:$};

\draw [edge] (v4) edge (v5);
\draw [Arrow] (v5) -- (v3);
\end{scope}

\begin{scope}[shift={(-0.4,-1.2)}]
\node [inode] (v2) at (-2.25,2.75) {};
\node [hidden] (v1) at (-2.8,3.4) {};
\node [hidden] (v3) at (-1.8,3.4) {};
\draw [Arrow] (v1) -- (v2);
\draw [Arrow] (v2) -- (v3);

\node at (-1.4,2.75) {$\rightarrow$};

\node [inode] (v4) at (-0.65,2.75) {};
\node [inode] (v5) at (0.35,2.75) {};
\node [hidden] (v3) at (0.8,3.4) {};
\node [hidden] (v1) at (-1.2,3.4) {};
\node [textnode] at (-2.96,2.74) {$2:$};

\draw [edge] (v4) edge (v5);
\draw [Arrow] (v5) -- (v3);
\draw [Arrow] (v1) -- (v4);
\end{scope}

\begin{scope}[shift={(-0.4,-2.4)}]
\node [inode] (v2) at (-2.25,2.75) {};
\node [hidden] (v1) at (-2.78,3.4) {};
\node [hidden] (v3) at (-1.8,3.4) {};
\draw [Arrow] (v1) -- (v2);
\draw [Arrow] (v2) -- (v3);

\node at (-1.4,2.75) {$\rightarrow$};

\node [inode] (v4) at (-0.65,2.75) {};
\node [inode] (v5) at (0.35,2.75) {};
\node [hidden] (v3) at (0.8,3.4) {};
\node [hidden] (v1) at (-0.2,3.4) {};
\node [textnode] at (-2.96,2.74) {$3:$};

\draw [edge] (v4) edge (v5);
\draw [Arrow] (v5) -- (v3);
\draw [Arrow] (v1) -- (v5);
\end{scope}

\begin{scope}[shift={(-0.4,-3.6)}]
\node [inode] (v2) at (-2.25,2.75) {};
\node [hidden] (v1) at (-2.8,3.4) {};
\node [hidden] (v3) at (-1.8,3.4) {};
\draw [Arrow] (v1) -- (v2);
\draw [Arrow] (v2) -- (v3);

\node at (-1.4,2.75) {$\rightarrow$};

\node [inode] (v4) at (-0.65,2.75) {};
\node [inode] (v5) at (0.35,2.75) {};
\node [hidden] (v3) at (0.82,3.4) {};
\node [hidden] (v1) at (-0.2,3.4) {};
\node [textnode] at (-2.96,2.74) {$4:$};
\node [hidden] (v6) at (-1.2,3.4) {};

\draw [edge] (v4) edge (v5);
\draw [Arrow] (v5) -- (v3);
\draw [Arrow] (v1) -- (v5);
\draw [Arrow] (v6) -- (v4);
\end{scope}

\begin{scope}[shift={(4.3,0)}]
\node [inode] (v2) at (-2.25,2.75) {};
\node [hidden] (v1) at (-2.86,3.4) {};
\node [hidden] (v3) at (-1.68,3.4) {};
\draw [Arrow] (v1) -- (v2);
\draw [Arrow] (v2) -- (v3);

\node at (-1.4,2.75) {$\rightarrow$};

\node [inode] (v4) at (-0.65,2.75) {};
\node [inode] (v5) at (0.35,2.75) {};
\node [hidden] (v3) at (0.8,3.4) {};
\node [textnode] at (-2.96,2.74) {$5:$};
\node [hidden] (v6) at (-1.2,3.4) {};
\node [hidden] (v7) at (-0.18,3.41) {};

\draw [edge] (v4) edge (v5);
\draw [Arrow] (v5) -- (v3);
\draw [Arrow] (v6) -- (v4);
\draw [Arrow] (v4) -- (v7);
\end{scope}

\begin{scope}[shift={(4.3,-1.2)}]
\node [inode] (v2) at (-2.25,2.75) {};
\node [hidden] (v1) at (-2.8,3.4) {};
\node [hidden] (v3) at (-1.7,3.4) {};
\draw [Arrow] (v1) -- (v2);
\draw [Arrow] (v2) -- (v3);

\node at (-1.4,2.75) {$\rightarrow$};

\node [inode] (v4) at (-0.65,2.75) {};
\node [inode] (v5) at (0.35,2.75) {};

\node [textnode] at (-2.96,2.74) {$6:$};
\node [hidden] (v6) at (-1.06,3.4) {};
\node [hidden] (v7) at (-0.32,3.42) {};
\draw [edge] (v4) edge (v5);

\draw [Arrow] (v6) -- (v4);
\draw [Arrow] (v4) -- (v7);
\end{scope}

\begin{scope}[shift={(4.3,-2.4)}]
\node [inode] (v2) at (-2.25,2.75) {};
\node [hidden] (v1) at (-2.8,3.4) {};
\node [hidden] (v3) at (-1.7,3.4) {};
\draw [Arrow] (v1) -- (v2);
\draw [Arrow] (v2) -- (v3);

\node at (-1.4,2.75) {$\rightarrow$};

\node [inode] (v4) at (-0.65,2.75) {};
\node [inode] (v5) at (0.35,2.75) {};
\node [hidden] (v3) at (-0.1,3.4) {};

\node [textnode] at (-2.96,2.74) {$7:$};

\node [hidden] (v7) at (-0.3,3.4) {};
\draw [edge] (v4) edge (v5);
\draw [Arrow] (v3) -- (v5);

\draw [Arrow] (v4) -- (v7);
\end{scope}

\begin{scope}[shift={(4.3,-3.6)}]
\node [inode] (v2) at (-2.25,2.75) {};
\node [hidden] (v1) at (-2.8,3.4) {};
\draw [Arrow] (v1) -- (v2);

\node at (-1.4,2.75) {$\rightarrow$};

\node [inode] (v4) at (-0.65,2.75) {};
\node [inode] (v5) at (0.35,2.75) {};
\node [textnode] at (-2.96,2.74) {$8:$};

\node [hidden] (v7) at (-1.08,3.42) {};
\draw [edge] (v4) edge (v5);
\draw [Arrow] (v7) -- (v4);
\end{scope}
\node [hidden] (a) at (0.8,3.1564) {};
\node [hidden] (b) at (0.8,-1) {};
\draw [dotted] (a) -- (b);

\end{tikzpicture}
    \caption{Set of all size 2 rules discovered from the initial example graph (shown in Figures.~\ref{fig:running_example} and \ref{fig:running_example}) Ultimately, BUGGE will choose to extract the third rule.}.
    \label{fig:all_initial_rules}
\end{figure}
}

\vspace{.2cm}
\noindent\textbf{Step 3: Finding the Rule with the Best Compression.} At this point, each connected subgraph is matched with one or more possible rules, and each matching may have a non-zero cost associated. The next step is to decide which rule should be extracted from the graph. For this, we revisit the MDL principle introduced in the previous section. Simply put, we select the rule that we predict will minimize the overall description length of the graph given the grammar. See Appendix~\ref{appendix-a} for details on how we encode graphs via grammar rules and measure them in bits.

We predict the number of bits it will cost to use a rule $n$ number of times as follows. Let $R$ be a rule with multiple occurrences in graph $H$ at various costs (number of edges to be added or deleted) $c_1, c_2, \ldots, c_m$, and let $x_i$ denote the number of time rules $R$ occurs in $H$ at cost $c_i$.

We extract the cheapest occurrences of the rule first (\ie, occurrences at cost $c_1$, followed by those at $c_2$, and so on.). Let $j(n)$ denote the highest cost index we would reach while extracting $n$ rule occurrences and $X_{j(n)} = n - \sum_{i=0}^{j-1}x_i$ be the number of rules at cost $c_j$ that we would select.

Let $C_R$ denote the cost to encode the rule itself and give it an identification number. $C_R$ will be zero if $R$ has already been extracted and encoded. Let $C_\text{ID}$ denote the cost to reference $R$'s ID number. Due to our encoding scheme, we only need to reference this id once to perform a series of extractions using the rule. Let $C_\text{node}$ be the cost to identify a single node in $H$ (the node that the rule would be applied to). Lastly, let $C_\text{edit}$ be the cost in bits to denote adding or deleting a node in $H$.

The predicted cost to encode $n$ occurrences of $R$ then becomes:
\begin{align}
    \text{COST}(n) =& \ C_R + C_\text{ID} + nC_\text{node} + X_{j(n)}c_{j(n)}C_\text{edit}  \nonumber \\ 
    & + \sum_{i=1}^{j(n)-1}x_ic_iC_\text{edit}
\end{align}

Recall that our MDL-based heuristic for selecting the most representative rule is to select the rule that lets us describe as much as possible with the fewest bits. Thus, what we really want to consider is not just the cost to encode some number $n$ of grammar rule extractions but rather the cost in bits \textit{per the number of nodes extracted}. We try to maximize the number of nodes per bit, which we refer to as the ``Predicted Cost Ratio'' (PCR). PCR for a rule is defined relative to the number of extractions that would yield the greatest ratio of nodes to bits.

Let $n_i$ represent the number of nodes in $H$ that would be extracted by a rule at a cost $c_i$. Thus, for a given $n$ extractions with a rule, the predicted number of nodes to be extracted would be:
\begin{equation}
    \text{NODES}(n) = \frac{X_{j(n)}}{x_{j(n)}}\, n_{j(n)} + \sum_{i=1}^{j(n)-1} n_i 
\end{equation}

The ideal predicted cost ratio (PCR) of nodes to bits for a rule $R$ then becomes:
\begin{equation}
    PCR_R = \max_n \frac{\text{NODES}(n)}{\text{COST}(n)}
\end{equation}

If BUGGE were to compute this directly, it would require checking every possible $n$ for each rule. Fortunately, it turns out that due to the ``overhead'' of the cost to encode and identify a rule, PCR will be maximized when \textit{all} the occurrences at a given cost are extracted. Thus, the calculation of PCR can be simplified to:
\begin{equation}
    PCR_R = \max_j \frac{\sum_{i=1}^j n_i}{C_R + C_\text{ID} + \sum_{i=1}^j x_i(C_\text{node} + c_iC_\text{edit})}
\end{equation}

We choose the rule with the highest PCR as the best rule to extract. Although we compute the best number of occurrences to extract when determining the PCR of a rule, this value may change as soon as a single extraction is performed because the changes in the graph may remove other occurrences of $R$, causing $R$ to have a worse PCR, or it causes some other rule $R^\prime$ to become cheaper or both. Hence, it should be stressed that this is a \textit{Predicted} Cost Ratio.

Also, note that BUGGE assumes that the sets of nodes covered by a rule at different cost levels are disjoint. This is an idealized assumption and could lead to inaccuracies, although the PCR ratio appears to performs well in practice.

Returning to our running example, we find that the rule with the best PCR is Rule 1 from Fig.~\ref{fig:rule_extract}(A). Although it has an occurrence at cost 1 in Fig.~\ref{fig:rule_extract}(A), this rule occurs three other times: twice with a cost of 0 and once more with a cost of 2. The extra occurrences at cost 0 are what give Rule 1 the best PCR. Thus BUGGE selects one of the cheapest occurrences of Rule 1 to extract (either \{a,b\} or \{e,d\}).

\vspace{.2cm}
\noindent\textbf{Step 4: Extracting a Rule Occurrence.} 
The rule extraction process ``collapses'' the induced subgraph by applying the rule in reverse. That is, instead of growing the graph by replacing a single vertex with a graph fragment as in Fig.~\ref{fig:rule}~and~\ref{fig:single-rule-example}, we reverse this process and extract a rule.

This processes is fairly straightforward. All the necessary edge additions or deletions were found when the rule occurrence was enumerated. Thus, BUGGE simply replaces the occurrence with a single node and rewires it according to the selected rule.

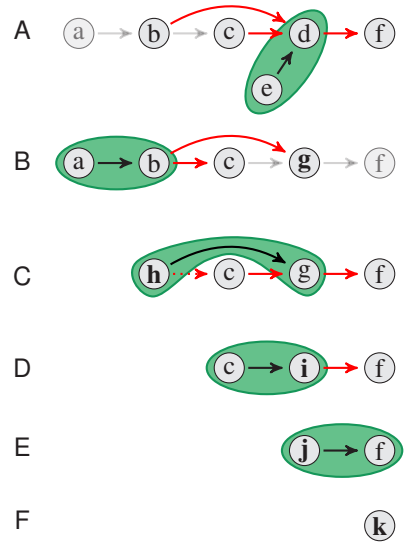
\begin{figure}
    \centering
    \begin{tikzpicture}[node distance=1.5cm, align=center, ]
\begin{scope}
\draw[ thick, draw=green!80!black, fill=green!60, rotate=55]  (3.6866,1.0472) ellipse (0.8 and 0.35);
\node [enode, faded] at (-1.5,4) (a) {a};
\node [enode] at (-0.5,4) (b) {b};
\node [enode] at (0.5,4) (c) {c};
\node [enode] at (1.5,4) (d) {d};
\node [enode] at (2.5,4) (f) {f};
\node [enode] at (1,3.25) (e) {e};

\draw[edge, faded, opacity=0.2] (a) -- (b);
\draw[edge, faded, opacity=0.2] (b) -- (c);
\draw[edge] (b) edge[bend left, draw=red] (d);
\draw[edge, draw=red] (c) -- (d);
\draw[edge] (e) -- (d);
\draw[edge, draw=red] (d) -- (f);
\end{scope}

\begin{scope}[shift={(0,-1.7248)}]
\draw[ thick, draw=green!80!black, fill=green!60]  (-1,4) ellipse (0.8 and 0.35);
\node [enode] (a) at (-1.5,4) {a};
\node [enode] (b) at (-0.5,4) {b};
\node [enode] (c) at (0.5,4) {c};
\node [enode] (g) at (1.5,4) {\textbf{g}};
\node [enode, faded] (f) at (2.5,4) {f};

\draw[edge] (a) -- (b);
\draw[edge, draw=red] (b) -- (c);
\draw[edge] (b) edge[bend left, draw=red] (g);
\draw[edge, opacity=0.2] (c) -- (g);
\draw[edge, opacity=0.2] (g) -- (f);
\end{scope}

\begin{scope}[shift={(0,-3.1998)}]
\fill[thick, fill=green!60, draw=green!80!black]  plot[smooth cycle, mark=none, tension=0.7, scale=0.25] coordinates {(2.1,17.95) (6.75,16.85) (6.15,14.85) (3.1,16.8) (0.85,16.75) (-2.05,14.8) (-2.85,16.7) };

\node [enode] (h) at (-0.5,4) {\textbf{h}};
\node [enode] (c) at (0.5,4) {c};
\node [enode] (g) at (1.5,4) {g};
\node [enode] (f) at (2.5,4) {f};

\draw[edge, draw=red, dotted] (h) -- (c);
\draw[edge] (h) edge[bend left, draw=black] (g);
\draw[edge, draw=red] (c) -- (g);
\draw[edge, draw=red] (g) -- (f);
\end{scope}

\begin{scope}[shift={(0,-4.45)}]
\draw[ thick, draw=green!80!black, fill=green!60]  (1,4) ellipse (0.8 and 0.35);

\node [enode] (c) at (0.5,4) {c};
\node [enode] (i) at (1.5,4) {\textbf{i}};
\node [enode] (f) at (2.5,4) {f};

\draw[edge] (c) -- (i);
\draw[edge, draw=red] (i) -- (f);
\end{scope}

\begin{scope}[shift={(0,-5.55)}]
\draw[ thick, draw=green!80!black, fill=green!60]  (2,4) ellipse (0.8 and 0.35);

\node [enode] (j) at (1.5,4) {\textbf{j}};
\node [enode] (f) at (2.5,4) {f};

\draw[edge] (j) -- (f);
\end{scope}

\begin{scope}[shift={(0,-6.55)}]
\node [enode] (k) at (2.5,4) {\textbf{k}};
\end{scope}

\node [textnode] at (-2.25,4) {\large\textsf{A}};
\node [textnode] at (-2.25,2.25) {\large\textsf{B}};
\node [textnode] at (-2.25,0.725) {\large\textsf{C}};
\node [textnode] at (-2.25,-0.5) {\large\textsf{D}};
\node [textnode] at (-2.25,-1.55) {\large\textsf{E}};
\node [textnode] at (-2.25,-2.525) {\large\textsf{F}};
\end{tikzpicture}
    \caption{BUGGE will repeatedly extract Rule 1 from Fig~\ref{fig:rule_extract}(A) thereby collapsing the entire example graph. Green areas highlight the nodes corresponding to a rule occurrence. Red arrows are boundary edges. The dotted red arrow in extraction (C) is an edge deletion. New vertices formed by the extraction of a rule (and the collapse of the relevant subgraph) in the previous step are labeled in bold. Note that the green highlighted nodes might lack in or out boundary edges. For example in (B), Rule 1 is extracted without cost from the subgraph a$\to$b despite no incoming edges to b. This is compatible with the grammar rule because KT-grammars require boundary edges {\em that exist} to be rewired according to the rule.}
    \label{fig:collapse}
\end{figure}

Returning to the running example, Fig.~\ref{fig:collapse} illustrates an extraction of Rule 1 where it is calculated to have the lowest cost.

\vspace{.2cm}
\noindent\textbf{Step 5: Update and Repeat.} An extraction changes the graph. So before we can iterate it is important that we update our record of rule occurrences.

To do this, we first determine which nodes have rule associations that may have changed due to the extraction in the previous step. Next, we delete registered rule occurrences involving any of the affected nodes. After the enumerations are updated, we repeat this process from Step 3. 

The set of nodes which might be affected is upper-bounded by the set of nodes connected to the subgraph that was extracted. More specifically, it is the union of the following sets:

\begin{itemize}
    \item The set containing the new ``collapsed'' node itself.
    \item Nodes in $H$ for which an edge was deleted or added in the process matching the rule.
    \item Nodes in $H$ which were connected by multiple in-edges or multiple out-edges to the collapsed subgraph, \ie, boundary edges.
\end{itemize}


Again we return to the running example in Fig~\ref{fig:collapse}. After Rule 1 is extracted from the pair $\{e,d\}$, updates to rule occurrence enumeration occur for any set involving the newly created node $g$. Of particular interest, after extracting Rule 1, $\{c, g\}$ (formerly $\{c, d\}$) ceases to have Rule 1 as one of its cheapest rules, but $\{b, g\}$ then has an occurrence of Rule 1 at cost 1, and it is eventually selected. During its run on the example graph, BUGGE extracts the entire graph using Rule 1 multiple times, albeit with a non-zero cost (Fig.~\ref{fig:collapse}~C). Recall that a boundary edge in a KT grammar rule indicates that ``all'' (0 or more) edges get wired to that node. Some of the extractions in our running example have no (in or out) boundary edges. 

\begin{figure*}[t!]
    \centering
    \begin{tikzpicture}
\begin{groupplot}[
    group style={
        group size=3 by 2,
        xlabels at=edge bottom,
        vertical sep=5pt,
        horizontal sep=5pt,
    },
    width  = 2.25in,
    height = 1.4 in,
    xlabel = {},
    scaled y ticks = false,
    axis line style={-},
    legend style={
    	cells={align=right}
    },
    x tick label style={
        /pgf/number format/.cd,
            precision=1,
        /tikz/.cd
    }
]

    \nextgroupplot[
    ylabel={\footnotesize\sffamily{Compression Rate}},
    ymin=-0.15, ymax=1.0,
    ylabel shift = 1 pt,
    xtick={},
    xticklabels={,,},
    ]
    \addplot+[opacity=0.6] coordinates {  
        (1000, 1-0.169)
        (2000, 1-0.164)
        (3000, 1-0.16)
        (4000, 1-0.16)
        (5000, 1-0.161)
    };
    
    \addplot+[opacity=0.6, dashed] coordinates {  
        (1000, 1-0.1978)
        (2000, 1-0.1858)
        (3000, 1-0.1825)
        (4000, 1-0.1816)
        (5000, 1-0.1812)
    };
    
    \addplot[opacity=0.6, dotted, green, mark=triangle*] coordinates {  
        (1000, 1-0.1987)
        (2000, 1-0.1916)
        (3000, 1-0.1880)
        (4000, 1-0.1876)
        (5000, 1-0.1858)
    };
    
    \nextgroupplot[
    xtick={},
    yticklabels={,,},
    xticklabels={,,},
    ]
    
    \addplot+[opacity=0.6,] coordinates {  
        (2, 1-0.942)
        (3, 1-0.467)
        (4, 1-0.315)
        (5, 1-0.317)
        (6, 1-0.199)
        (7, 1-0.160)
        (8, 1-0.151)
    };
    
    \addplot+[opacity=0.6, dashed] coordinates {  
        (2, 1-0.879)
        (3, 1-0.446)
        (4, 1-0.296)
        (5, 1-0.2326)
        (6, 1-0.182)
        (7, 1-0.178)
        (8, 1-0.141)
    };
    
    \addplot[opacity=0.6, dotted, green, mark=triangle*] coordinates {  
        (2, 1-1.038)
        (3, 1-0.538)
        (4, 1-0.359)
        (5, 1-0.279)
        (6, 1-0.225)
        (7, 1-0.188)
        (8, 1-0.163)
    };
    
    \nextgroupplot[
    yticklabels={,,},
    xticklabels={,,},
    xmode=log,
    xtick={0.001,0.01,0.1,0.99},
    ]
    
    \addplot+[opacity=0.6] coordinates {  
        (0.001, 1-0.1600571293)
        (0.0025, 1-0.1692216139)
        (0.005, 1-0.1876934063)
        (0.01, 1-0.2109735777)
        (0.02, 1-0.2406807903)
        (0.04, 1-0.3129016901)
        (0.08, 1-0.3667460129)
        (0.16, 1-0.48907403)
        (0.32, 1-0.6204475125)
        (0.64, 1-0.6590573673)
        (1, 1-0.710259462)
    };
    
    \addplot+[opacity=0.6, dashed] coordinates {  
        (0.001, 1-0.1824927146)
        (0.0025, 1-0.2010984084)
        (0.005, 1-0.2122394082)
        (0.01, 1-0.2454382425)
        (0.02, 1-0.2952925353)
        (0.04, 1-0.3811701412)
        (0.08, 1-0.4287379511)
        (0.16, 1-0.5237614885)
        (0.32, 1-0.6275274602)
        (0.64, 1-0.6988119256)
        (1, 1-0.7101322573)
    };
    
    \addplot[opacity=0.6, dotted, green, mark=triangle*] coordinates {  
        (0, 1-0.1880330277)
        (0.0025, 1-0.2470779902)
        (0.005, 1-0.2559643558)
        (0.01, 1-0.2815003147)
        (0.02, 1-0.3217234612)
        (0.04, 1-0.4913419646)
        (0.08, 1-0.6898041297)
        (0.16, 1-0.8889697987)
        (0.32, 1-0.9960628464)
        (0.64, 1-1.103674265)
        (1, 1-1.065697395)
    };
    
    \nextgroupplot[
    ylabel={\footnotesize\sffamily{Runtime (log s)}},
    xlabel={\footnotesize\sffamily$|V|$},
    ymin=1, ymax=1000000,
    xtick={1000,2000,3000,4000,5000}, 
    ytick = {1,100,10000,1000000},
    ymode=log,
    ]
    \addplot+[opacity=0.6] coordinates {  
        (1000, 138.104)
        (2000, 695.326)
        (3000, 284.472)
        (4000, 5481.128)
        (5000, 546.059)
    };
    
    \addplot+[opacity=0.6, dashed] coordinates {  
        (1000, 181.507)
        (2000, 351.845)
        (3000, 516.323)
        (4000, 710.985)
        (5000, 895.96)
    };
    
    \addplot[opacity=0.6, dotted, green, mark=triangle*] coordinates {  
        (1000, 274.707)
        (2000, 538.105)
        (3000, 1071.641)
        (4000, 1060.701)
        (5000, 1341.575)
    };
    
    \nextgroupplot[
    symbolic x coords={2,3,4,5,6,7,8}, 
    xtick=data,
    ymin=1, ymax=1000000,
    xlabel={\footnotesize\sffamily{Rule Max Size}},
    yticklabels={,,},
    ymode=log
    ]
    
    \addplot+[opacity=0.6, ] coordinates {  
        (2, 8.514)
        (3, 23.56)
        (4, 33.103)
        (5, 92.735)
        (6, 167.511)
        (7, 284.472)
        (8, 9448.861)
    };
    
    \addplot+[opacity=0.6, dashed] coordinates {  
        (2, 11.383)
        (3, 41.426)
        (4, 107.126)
        (5, 239.533)
        (6, 516.323)
        (7, 2386.096)
        (8, 84142.435)
    };
    
    \addplot[opacity=0.6, dotted, green, mark=triangle*] coordinates {  
        (2, 43.847)
        (3, 92.775)
        (4, 221.627)
        (5, 387.101)
        (6, 603.991)
        (7, 1076.811)
        (8, 1120.429)
    };
    
    \nextgroupplot[
    xticklabels={0,0.01,0.1,1}, 
    xtick={0.001,0.01,0.1,0.99},
    xmax=1.4,
    ymin=1, ymax=1000000,
    xlabel={\footnotesize\sffamily{Rewiring Probability (log)}},
    yticklabels={,,},
    ymode=log,
    xmode=log,
    ]    
    
    \addplot+[opacity=0.6, ] coordinates {  
        (0.001, 284.472)
        (0.0025, 996.885) 
        (0.005, 7675.94)
        (0.01, 12419.34)
        (0.02, 18760.846)
        (0.04, 141468.141)
        (0.08, 193188.446)
        (0.16, 258705.447)
        (0.32, 191681.811)
        (0.64, 288478.094)
        (1, 134239.468)
    };
    
    \addplot+[opacity=0.6, dashed] coordinates {  
        (0.001, 516.323)
        (0.0025, 4928.068)
        (0.005, 4889.255)
        (0.01, 4763.292)
        (0.02, 43409.529)
        (0.04, 46009.578)
        (0.08, 100898.17)
        (0.16, 125697.536)
        (0.32, 117288.739)
        (0.64, 110936.799)
        (1, 136065.858)
    };
    
    \addplot[opacity=0.6, dotted, green, mark=triangle*] coordinates {  
        (0.001, 1071.641)
        (0.0025, 6670.213)
        (0.005, 1607.282)
        (0.01, 1499.766)
        (0.02, 1835.681)
        (0.04, 11108.883)
        (0.08, 7093.248)
        (0.16, 12967.158)
        (0.32, 14179.592)
        (0.64, 10434.132)
        (1, 10546.263)
    };
    
\end{groupplot}
\end{tikzpicture}
    \begin{tikzpicture}
    \begin{customlegend}[ 
    legend columns=3,
    legend style={
    draw=none,
    font=\sffamily\footnotesize,
    column sep=2ex,
  },
  legend entries={Binary Tree, Tree of Rings, Ring Lattice},
  ]
    \addlegendimage{blue, mark=*, opacity=0.6}
    \addlegendimage{red, dashed, mark=square*, opacity=0.6}
    \addlegendimage{green, dotted, mark=triangle*, opacity=0.6, }
    \end{customlegend}
\end{tikzpicture}
    \caption{BUGGE's Compression and Runtime results for synthetic graphs. We see how BUGGE responds as we vary different parameters: the size of the graph, the maximum allowed grammar rule size, and the amount of noise in the input (i.e. rewiring probability). 
    }
    \label{fig:synthetic_r_and_c}
    \vspace{-.5cm}
\end{figure*}
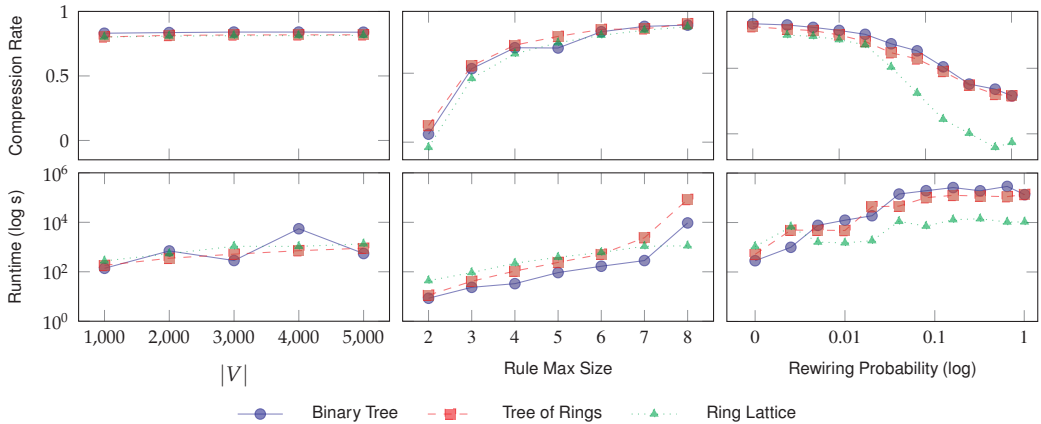

\vspace{.2cm}
\noindent\textbf{Enumeration Heuristic.} We found during testing that in practice, the rules with the cheapest edit costs are usually the rules with the best Predicted Compression Ratio (PCR). Thus, during the process of enumerating rules, it would only be important to enumerate the rules which have the lowest or near-lowest edit costs.

Consider a connected set $X$ of $k < k_\text{max}$ nodes with an edit cost of $c$. This means that there are $c$ edges which must be added and/or deleted in order to extract $X$ into a rule. The only way that adding another node to $X$ could reduce the cost is if that node is one of the nodes that are connected to $X$ via one of the edges which must be added or deleted. Furthermore, this new node must not add any more edit cost. Thus, the chances of the cost decreasing as nodes are added are usually quite low.

BUGGE takes advantage of this observation. BUGGE stores the cost of the cheapest rule occurrence $c_\text{best}$; then, during rule occurrence enumeration, it updates this cost. If during enumeration, it finds that a set of nodes $X$ has a cost which is ``too far'' from $c_\text{best}$ then it doesn't bother to enumerate any connected sets of which $X$ is a subset. We find that this provides a dramatic speedup.

More formally, we define a ``shortcut parameter'' $s$ which tells BUGGE whether or not to enumerate larger sets. Specifically, we continue enumerating supersets of a set $X$ with edit cost $c$ ($X = k < k_\text{max}$) if the following inequality holds:
\begin{equation}
    c \leq c_\text{best} + \min \left\{1 + k_\text{max} - k,\, s + \lceil \ln(k_\text{max} - k) \rceil)\right\}
\end{equation}

In practice, we find that setting the shortcut parameter $s$ to 1 tends to produce very similar results to running without a shortcut at all yet with a drastically reduced runtime (particularly noticeable when $k_\text{max}$ is large). Larger values may increase runtime but produce better results. Sometimes we find that $s = 2$ will find interesting results that $s = 1$ will not; thus, $s$ should be treated as a parameter that allows a potential tradeoff between results quality and runtime. Even with larger values of $s$ however, the runtime is usually significantly reduced.

\subsection{Related Work}\label{sec:related_work}

\vspace{.2cm}
\noindent\textbf{MDL Approaches} Of other well-known graph mining systems, our approach is most akin to SUBDUE \cite{holder1994substucture}, followed by VoG (Vocabulary of Graphs) \cite{koutra2014vog}. Like our system, both SUBDUE and VoG use the MDL heuristic to select structures to extract. 

SUBDUE uses a beam search to find structures. VoG searches for 6 preset structure types (cliques, stars, etc.) and maybe extended if the user wishes to implement support for other specific structures. Our system finds whatever structures are present in the graph up to a user-specified number of nodes. Thus, our approach lets the data ``speak for itself'' up to whatever computational costs the user is willing to allow.


\begin{figure*}[t!]
    \centering
    \input{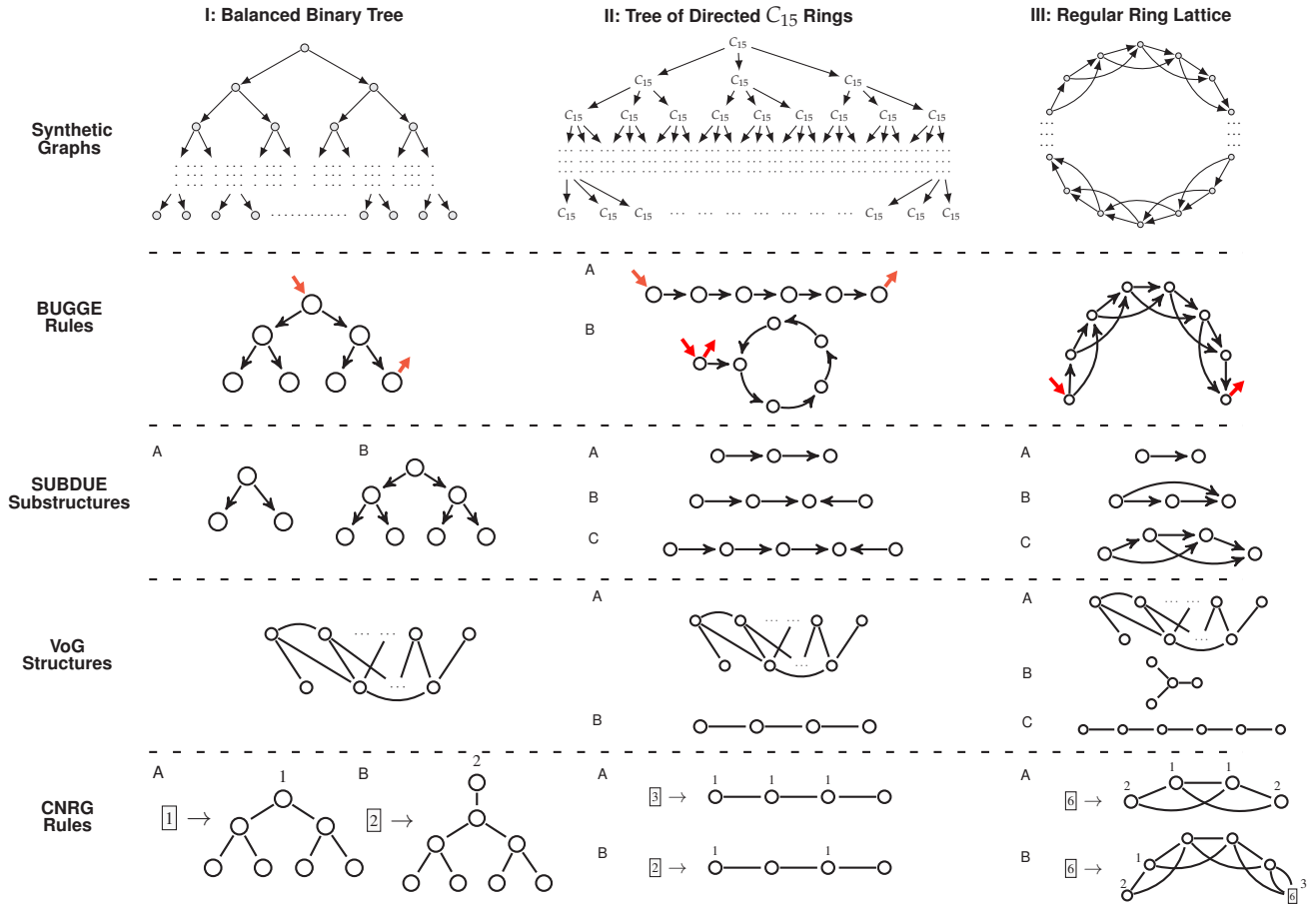}
    \caption{Rules or substructures extracted by graph mining for three types of synthetic graphs. The rules extracted by BUGGE capture the known dynamics of the synthetic graph. 
    }
    \label{fig:synthetic_extractions}
    \vspace{-.5cm}
\end{figure*}

\vspace{.2cm}
\noindent\textbf{Graph Grammar Approaches} Other approaches extract a vertex replacement grammar using either a hierarchical clustering \cite{sikdar2019modeling} or a tree decomposition of a graph \cite{aguinaga2018learning} to select which nodes to collapse into a grammar rule. These approaches effectively try to form grammar rules from nodes that ``go together." The Clustering-based Node Replacement Grammar (CNRG) provides a computational advantage over our approach. However, the choice of clustering algorithm adds a layer of indirection between the graph and the grammar which detracts from compressibility and interpretability.

The original work of Kemp and Tennenbaum did not extract grammars from a graph but instead tested if a dataset matched a particular grammar rule. This was particularly insightful because KT-grammars have two particular advantages. 

First KT-grammars tend to be naturally interpretable when the intelligible structure can be found. Of particular noteworthiness, these rules can easily capture many of the structures which are most intuitive to the human mind: trees, rings, hierarchies, etc. For example, Table \ref{fig:synthetic_extractions} shows some graphs along with grammar rules which generate them.

Second, KT-grammars are robust to error. For a given subgraph, there might not exist a rule that can create that particular subgraph. Usually, this happens when two nodes with external outgoing (or incoming) edges do not all point to the same nodes. At first glance, this might seem to be a weakness, but it enables an intuitive notion of the ``cost'' of applying a rule to a subgraph. This cost, defined as the number of edges in the graph that need to be added and/or deleted before the rule could apply, allows our algorithm to focus on the parts of the graph that most clearly correspond to interpretable structure, compressing those parts of the graph first.

\vspace{.2cm}
\noindent\textbf{Other Approaches} Exponential Random Graph Models (ERGMs) are another type of graph model that learns a robust graph model from user-defined features of a graph~\cite{robins2007introduction}. Unfortunately, this model does not scale well and is prone to model degeneracy. 
Neural network graph models are of recent interest, but as is common with neural networks, these models do not provide the interpretability we desire. Additionally, some, such as GraphVAE~\cite{simonovsky2018graphvae} and GraphRNN~\cite{you2018graphrnn} have limited scalability while others such as NetGAN~\cite{bojchevski2018netgan} produce models many times larger than the original graph. Node embedding models like LINE~\cite{tang2015line}, node2vec~\cite{grover2016node2vec}, VGAE~\cite{kipf2016variational}, and others~\cite{goyal2018graph} represent individual nodes in the context of their local substructures for classification or prediction tasks and are also poorly suited to our objective.


\section{Methodology}

In this section we present results of extensive experiments on real and synthetic datasets that compare compression, runtime, and model interpretability. We compare our results to several state of the art graph summarization and grammar extraction methods including VoG~\cite{koutra2014vog}, SUBDUE~\cite{holder1994substucture}, and CNRG~\cite{sikdar2019modeling}. The source code for BUGGE, including experimental data and evaluation scripts, is available on GitHub\footnote{\url{https://github.com/SteveWillowby/ThreePartsTree}}.

\vspace{.2cm}
\noindent\textbf{Datasets.} It is important that we consider both synthetic and real-world graphs in our evaluation. Synthetic graphs enable us to determine whether or not the grammar rules that we extract are interpretable, \ie, since we know how we generate some synthetic graph, it's relatively easy to determine the goodness of the found graph substructures.

To that end, we generate three types of synthetic graphs: (1) a Binary Tree, (2) a Tree of Rings, which is an $N$-ary tree where each node is replaced with a ring of size $k$, and (3) a Ring Lattice, based on the Watts-Strogatz model of social networks. 
An ideal grammar extractor would describe these simple structures clearly.

In addition, we consider three real-world directed graphs from SNAP: Blogs (1,224 nodes, 19,025 edges), Protein-to-protein interaction network (1,706 nodes, 6,207 edges), and the DBLP citation graph (12,591 nodes, 49,743 edges). 

\nop{
\vspace{.2cm}
\noindent\textbf{Performance Metrics.} We measure performance on compression, runtime, interpretability. If BUGGE is able to compress a graph, that suggests that it has found some meaningful structure within. Runtime is important for practical considerations. Lastly, and perhaps most importantly, our manual examination of extracted grammar rules shows that BUGGE finds results which make \textit{interpretable} sense.

}

Because runtime drastically increases with the maximum rule size, we use the enumeration heuristic in all of our tests. We find that it preserves the quality of results while improving runtime dramatically.

\subsection*{Synthetic Graph Results}

First we test the runtime and compression ratio of BUGGE in various scenarios.  
Unless otherwise specified, graphs are generated with 3000 nodes; the $N$-ary tree has $n=3$ with ring size $k=15$; and the directed ring lattice graph has the degree set to 4.

\vspace{.2cm}
\noindent\textbf{Graph Size.} Holding the rule size steady and the rewiring probability at 0\%, we vary the number of nodes in the synthetic graph from 1000 to 5000. The results shown in Fig.~\ref{fig:synthetic_r_and_c}(left) illustrates that runtime increases linearly in the number of nodes. This is what we expect given that the larger synthetic graphs just have more repetition of the same structure, so for a fixed $k_\text{max}$ BUGGE just enumerates the same grammar rules more times. 

The compression rate improves slightly on larger graphs. This matches our expectation because the overhead of defining more rules (bits increase) in larger graphs is dwarfed by the number of extractions with that rule (bits savings).


\vspace{.2cm}
\noindent\textbf{Rule Size.} Holding the graph size and rewiring probability steady at 3000 and 0.0\% respectively, we vary the maximum rule size from 2 to 8. Size-2 rules correspond to edges; size-3 rules can be one of the 5 directed 3-node graphlets. There are 34 different size-4 directed graphlets, and this number increases dramatically as the maximum allowed rule size increases~\cite{sarajlic2016graphlet, milo2002network}. This increase in expressibility is certain to cause an increase in runtime. The results shown in Fig.~\ref{fig:synthetic_r_and_c}(center) illustrates that the compression rate increases dramatically as the rule size increases (higher is better). 

Model interpretability is explored in Fig.~\ref{fig:synthetic_extractions}, which illustrates the most frequent rules extracted by BUGGE and the comparison methods where parameters are set empirically for each dataset. For example, in the synthetic binary tree graph Fig.~\ref{fig:synthetic_extractions}(left), BUGGE extracts only two grammar rules, one of which is (re-)used in 499 of the 501 total iterations. Thus, almost the entire graph can be represented with a single rule, which corresponds to replacing a node with a subtree. SUBDUE and CNRG extract reasonable rules from the binary tree, but VoG surprisingly extracts a nearly bipartite core.

For the tree of rings Fig.~\ref{fig:synthetic_extractions}(center) illustrates two rules extracted by BUGGE that account for almost the entire graph (599 of the 601 extractions). First, a rule for a chain is used twice per ring to shrink the rings. Then a second rule takes one of the shrunken child-rings and wraps it entirely into its parent-ring. SUBDUE, VoG, and CNRG extract rules and substructures which are difficult to interpret.

For the ring lattice graph in Fig.~\ref{fig:synthetic_extractions}(right), BUGGE extracts an intuitive rule that comprises 427 of the 430 total extractions. SUBDUE and CNRG produce reasonable results; however dozens of other CNRG rules are not illustrated here, and SUBDUE's best rule accounts for at most 68\% of the graph. Again VoG produces a bipartite core.


These examples demonstrate how BUGGE can discern the nature of the original graph and common patterns within.

\vspace{.2cm}
\noindent\textbf{Random Rewriting Probability.} To test the robustness of BUGGE to noise, we define a rewiring probability $r$. Holding the graph size at 3000, we vary $r$ from 0 to 1. Before extraction, every edge is randomly re-assigned to a new pair of nodes with probability $r$. We design this process to ensure that the number of edges is preserved. This means that when $r$ is 0 the synthetic graph remains the same and when $r$ is 1 it becomes an Erdos Renyi graph.

Returning to Fig.~\ref{fig:synthetic_r_and_c}(right), we observe that runtime increases significantly as the level of noise increases and compressibility drops. Interestingly, BUGGE manages to compress the graph with increasing levels of noise, thereby showing robustness; even when the rewiring probability is 1 (\ie, entirely noise) BUGGE still manages to compress the two sparser random graphs, indicating that BUGGE can compress sparse noise.

\subsection*{Real-World Graph Results}
The synthetic graph results show that BUGGE does indeed extract grammar rules that are meaningful. By inspecting the rules, we can discern certain aspects about how the graph is structured. Real-world graphs are less straightforward, but the goal remains the same: to extract meaningful rules that describe the underlying structure of the graph. Ideally, these rules will hint at the dynamics of the graph and shed light on the processes that govern these large, complex systems. We extracted grammars from 3 real-world graphs and inspected them to see what they tell us about the original graphs's structure.






\begin{figure}
    \centering
    \begin{tikzpicture}
\begin{scope}
    \begin{axis}[
        title style={yshift=-1.75ex,},
        title={\footnotesize\sffamily{PPI}},
        axis y line*=left,
        width=7.5cm,
        height = 3.5cm,
        xticklabels={,,},
        ybar,
        bar width=12pt,
        major x tick style = transparent,
        ylabel = {\footnotesize\sffamily{Frequency}},
        xlabel = {},
        xmin=0.8, xmax=3.7,
        ymax=0.75,
        ymin=-0.01,
    ]
    \addplot[thick, blue,fill=blue!30,mark=none]
    coordinates {
       (1, 844/1177) 
       (1.5, 73/1177)
       (2, 60/1177)
       (2.5, 53/1177)
       (3, 36/1177)
       (3.5, 25/1177)
    };
    \end{axis}
    
    \begin{axis}[
    axis y line*=right,
    axis x line=none,
    width = 7.5cm,
    height = 3.5cm,
    xticklabels=\empty,
    major x tick style = transparent,
    axis line style={-},
    ylabel = {\footnotesize\sffamily{Edit Cost per Node}},
    ymin=-0.01, ymax=1350,
    xmin=0.8, xmax=3.7, 
    xtick=data,
    ]
    
    \addplot[thick, red, mark=*, mark size=1.5]
     coordinates {
     (1, 1280.5) 
     (1.5, 39.5) 
     (2, 0) 
     (2.5, 0) 
     (3, 0)
     (3.5, 0)
     };
    \end{axis}
\end{scope}

\begin{scope}[shift={(0.8,-1.67)}, scale=0.45, transform shape]  
\node [inode] (a) at (-1.025,2.75) {};
\node [inode] (b) at (-1,1.2006) {};
\node [hidden] (in) at (-1.6082,3.6332) {};
\node [hidden] (out) at (-0.5079,3.6432) {};

\draw [edge, <->] (a) edge (b);
\draw [Arrow] (in) -- (a);
\draw [Arrow] (a) -- (out);
\end{scope}

\begin{scope}[shift={(2.28,-1.69)}, scale=0.45, transform shape]  
\node [inode] (a) at (-2,2.75) {};
\node [inode] (c) at (-2.5549,1.4222) {};
\node [inode] (b) at (-1.089,1.6223) {};
\node [hidden] (in) at (-2.4999,3.6749) {};
\node [hidden] (out) at (-1.5112,3.6499) {};

\draw [edge, <->] (a) edge (b);
\draw [edge, <->] (b) edge (c);

\draw [Arrow] (in) -- (a);
\draw [Arrow] (a) -- (out);
\end{scope}

\begin{scope}[shift={(2.88,-1.71)}, scale=0.45, transform shape]  
\node [inode] (a) at (-1,2.75) {};
\node [inode] (b) at (-1,1.3338) {};
\node [hidden] (in) at (-1.5638,3.6332) {};
\node [hidden] (out) at (-0.5523,3.6432) {};

\draw [edge, ->] (a) edge (b);
\draw [Arrow] (in) -- (a);
\draw [Arrow] (a) -- (out);
\end{scope}

\begin{scope}[shift={(3.82,-1.71)}, scale=0.45, transform shape]  
\node [inode] (a) at (-0.8918,2.75) {};
\node [inode] (c) at (-0.3443,1.4783) {};
\node [inode] (b) at (-1.4329,1.4446) {};
\node [hidden] (in) at (-1.4306,3.6332) {};
\node [hidden] (out) at (-0.308,3.5988) {};

\draw [edge, <->] (a) edge (b);
\draw [edge, <->] (c) edge (a);

\draw [Arrow] (in) -- (a);
\draw [Arrow] (a) -- (out);
\end{scope}

\begin{scope}[shift={(4.8225,-1.62)}, scale=0.42, transform shape]  
\node [inode] (a) at (-0.85,2.75) {};
\node [inode] (b) at (0.3848,1.829) {};
\node [hidden] (in) at (-1.4166,3.6642) {};
\node [hidden] (out) at (-0.2908,3.6742) {};
\node [inode] (c) at (0.6322,2.512) {};
\node [inode] (d) at (-0.1968,1.345) {};
\node [inode] (e) at (-2.122,1.8528) {};
\node [inode] (f) at (-2.3094,2.5834) {};
\node [inode] (g) at (-1.2662,1.4164) {};

\draw [Arrow] (in) -- (a);
\draw [Arrow] (a) -- (out);
\draw [edge, <->] (a) edge (b);
\draw [edge, <->] (a) edge (c);
\draw [edge, <->] (a) edge (d);
\draw [edge, <->] (a) edge (e);
\draw [edge, <->] (a) edge (f);
\draw [edge, <->] (a) edge (g);
\end{scope}

\begin{scope}[shift={(5.93,-1.72)}, scale=0.45, transform shape]  
\node [inode] (a) at (-1,2.75) {};
\node [inode] (b) at (-0.3776,1.4225) {};
\node [hidden] (in) at (-1.5499,3.6249) {};
\node [hidden] (out) at (-0.4968,3.5905) {};
\node [inode] (c) at (0.1112,2.0004) {};
\node [inode] (d) at (-1.2222,1.3339) {};

\draw [Arrow] (in) -- (a);
\draw [Arrow] (a) -- (out);
\draw [edge, <->] (a) edge (b);
\draw [edge, <->] (a) edge (c);
\draw [edge, <->] (a) edge (d);
\end{scope}

\end{tikzpicture}
    \begin{tikzpicture}
\begin{scope}
    \begin{axis}[
        title style={yshift=-2.0ex,},
        axis y line*=left,
        width=7.5cm,
        height = 3.5cm,
        title style={yshift=0.25ex,},
        title={\footnotesize\sffamily{DBLP}},
        xticklabels={,,},
        ybar,
        bar width=12pt,
        major x tick style = transparent,
        ylabel = {\footnotesize\sffamily{Frequency}},
        xlabel = {},
        xmin=0.8, xmax=5.2,
        ymax=0.75,
        ymin=-0.0,
    ]
    \addplot[thick, blue,fill=blue!30,mark=none]
    coordinates {
       (1, 2667/8257) 
       (1.5, 1935/8257)
       (2, 1157/8257)
       (2.5, 855/8257)
       (3, 446/8257)
       (3.5, 430/8257)
       (4, 362/8257)
       (4.5, 240/8257)
       (5, 165/8257)
    };
    \end{axis}
    
    \begin{axis}[
    axis y line*=right,
    axis x line=none,
    width = 7.5cm,
    height = 3.5cm,
    xticklabels=\empty,
    major x tick style = transparent,
    axis line style={-},
    ylabel = {\footnotesize\sffamily{Edit Cost per Node}},
    ymin=-0.0, ymax=6000,
    xmin=0.8, xmax=5.2, 
    xtick=data,
    ]
    
    \addplot[thick, red,mark=*, mark size=1.5]
     coordinates {
     (1, 2502) 
     (1.5, 3511) 
     (2, 5795) 
     (2.5, 39) 
     (3, 0)
     (3.5, 59.33)
     (4, 51.66)
     (4.5, 1230.5)
     (5, 791)
     };
    \end{axis}
\end{scope}

\begin{scope}[shift={(1.3828,-1.668)}, scale=0.45, transform shape]  
\node [inode] (a) at (-1.025,2.75) {};
\node [inode] (b) at (-1,1.2156) {};
\node [hidden] (in) at (-1.6688,1.9767) {};
\node [hidden] (out) at (-0.6431,3.6583) {};

\draw [edge] (a) edge (b);
\draw [Arrow] (in) -- (b);
\draw [Arrow] (a) -- (out);
\end{scope}

\begin{scope}[shift={(4.7796,-1.662)}, scale=0.45, transform shape]  
\node [inode] (a) at (-1.025,2.75) {};
\node [inode] (b) at (-0.9772,1.1962) {};
\node [inode] (c) at (-0.375,1.7386) {};
\node [hidden] (in) at (-1.5022,2.0715) {};
\node [hidden] (out) at (-0.4993,3.5603) {};

\draw [edge] (a) edge (b);
\draw [edge] (a) edge (c);
\draw [Arrow] (in) -- (b);
\draw [Arrow] (a) -- (out);
\end{scope}

\begin{scope}[shift={(3.3212,-1.6832)}, scale=0.45, transform shape]  
\node [inode] (a) at (-0.7636,2.75) {};
\node [inode] (b) at (-0.0946,1.7223) {};
\node [inode] (c) at (-0.6472,1.1864) {};
\node [inode] (d) at (-1.1537,1.7619) {};
\node [hidden] (in) at (-0.3856,2.7419) {};
\node [hidden] (out) at (-0.1334,3.6583) {};

\draw [edge] (a) edge (b);
\draw [edge] (a) edge (c);
\draw [edge] (a) edge (d);
\draw [Arrow] (in) -- (b);
\draw [Arrow] (a) -- (out);
\end{scope}

\begin{scope}[shift={(0.6648,-1.668)}, scale=0.45, transform shape]  
\node [inode] (a) at (-1.025,2.75) {};
\node [inode] (b) at (-1,1.1864) {};
\node [hidden] (in) at (-1.525,3.6602) {};
\node [hidden] (out) at (-0.5124,3.6452) {};

\draw [edge] (a) edge (b);
\draw [Arrow] (in) -- (a);
\draw [Arrow] (a) -- (out);
\end{scope}

\begin{scope}[shift={(4.1032,-1.712)}, scale=0.45, transform shape]  
\node [inode] (a) at (-1.025,2.8807) {};
\node [inode] (b) at (-1,1.2418) {};
\node [inode] (c) at (-0.2443,1.6079) {};
\node [hidden] (in) at (-1.6057,3.706) {};
\node [hidden] (out) at (-0.4797,3.691) {};

\draw [edge] (a) edge (b);
\draw [edge] (a) edge (c);
\draw [Arrow] (in) -- (a);
\draw [Arrow] (a) -- (out);
\end{scope}

\begin{scope}[shift={(2.7236,-1.712)}, scale=0.45, transform shape]  
\node [inode] (a) at (-1.025,2.8807) {};
\node [inode] (b) at (-1.6455,1.7648) {};
\node [inode] (c) at (-1,1.2778) {};
\node [inode] (d) at (-0.3807,1.7386) {};
\node [hidden] (in) at (-1.4869,3.804) {};
\node [hidden] (out) at (-0.2641,3.6583) {};

\draw [edge] (a) edge (b);
\draw [edge] (a) edge (c);
\draw [edge] (a) edge (d);
\draw [Arrow] (in) -- (a);
\draw [Arrow] (a) -- (out);
\end{scope}

\begin{scope}[shift={(2.0384,-1.668)}, scale=0.45, transform shape]  
\node [inode] (a) at (-1.025,2.75) {};
\node [inode] (b) at (-1,1.1864) {};
\node [hidden] (in) at (-1.5,3.6602) {};
\node [hidden] (out) at (-0.4374,2.0859) {};

\draw [edge] (a) edge (b);
\draw [Arrow] (in) -- (a);
\draw [Arrow] (b) -- (out);
\end{scope}

\begin{scope}[shift={(5.4132,-1.662)}, scale=0.45, transform shape]  
\node [inode] (a) at (-1.025,2.75) {};
\node [inode] (b) at (-1,1.1995) {};
\node [hidden] (in) at (-1.6176,3.5491) {};
\node [hidden] (out) at (-0.4309,2.0794) {};

\draw [edge, <->] (a) edge (b);
\draw [Arrow] (in) -- (a);
\draw [Arrow] (b) -- (out);
\end{scope}

\begin{scope}[shift={(6.1708,-1.824)}, scale=0.5, transform shape]  
\node [inode] (a) at (-1.025,2.75) {};
\node [inode] (b) at (-1,1.306) {};
\node [hidden] (in) at (-1.6426,3.5426) {};
\node [hidden] (out) at (-0.5124,3.5452) {};

\draw [edge, <->] (a) edge (b);
\draw [Arrow] (in) -- (a);
\draw [Arrow] (a) -- (out);
\end{scope}
\end{tikzpicture}
    \begin{tikzpicture}
\begin{scope}
    \begin{axis}[
        axis y line*=left,
        title style={yshift=-2.0ex,},
        width=7.5cm,
        height = 3.5cm,
        title={\footnotesize\sffamily{PolBlogs}},
        xticklabels={,,},
        ybar,
        bar width=12pt,
        major x tick style = transparent,
        ylabel = {\footnotesize\sffamily{Frequency}},
        xlabel = {},
        xmin=0.8, xmax=3.7,
        ymax=0.75,
        ymin=-0.0,
    ]
    \addplot[thick, blue,fill=blue!30,mark=none]
    coordinates {
       (1, 323/1018) 
       (1.5, 226/1018)
       (2, 155/1018)
       (2.5, 127/1018)
       (3, 99/1018)
       (3.5, 88/1018)
    };
    \end{axis}
    
    \begin{axis}[
    axis y line*=right,
    axis x line=none,
    width = 7.5cm,
    height = 3.5cm,
    xticklabels=\empty,
    major x tick style = transparent,
    axis line style={-},
    ylabel = {\footnotesize\sffamily{Edit Cost per Node}},
    ymin=-0.0, ymax=2150,
    xmin=0.8, xmax=3.7, 
    xtick=data,
    ]
    
    \addplot[thick, red,mark=*, mark size=1.5]
     coordinates {
     (1, 2029) 
     (1.5, 478.5) 
     (2, 798.5) 
     (2.5, 217) 
     (3, 1225)
     (3.5, 1098.5)
     };
    \end{axis}
\end{scope}

\begin{scope}[shift={(0.8764,-1.78)}, scale=0.5, transform shape]  
\node [inode] (a) at (-1.025,2.75) {};
\node [inode] (b) at (-1,1.3) {};
\node [hidden] (out) at (-0.4588,3.5338) {};
\node [hidden] (in) at (-1.5962,2.0688) {};

\draw [edge,] (a) edge (b);
\draw [Arrow] (a) -- (out);
\draw [Arrow] (in) -- (b);
\end{scope}

\begin{scope}[shift={(4,-1.8032)}, scale=0.5, transform shape]  
\node [inode] (a) at (-1.025,2.75) {};
\node [inode] (b) at (-1,1.3588) {};
\node [hidden] (in) at (-1.525,3.525) {};
\node [hidden] (out) at (-0.4712,3.4688) {};

\draw [edge] (a) edge (b);
\draw [Arrow] (in) -- (a);
\draw [Arrow] (a) -- (out);
\end{scope}

\begin{scope}[shift={(2.9764,-1.8032)}, scale=0.5, transform shape]  
\node [inode] (a) at (-1.025,2.75) {};
\node [inode] (b) at (-1,1.3588) {};
\node [hidden] (in) at (-1.525,3.525) {};
\node [hidden] (out) at (-0.5712,3.5688) {};

\draw [edge, <->] (b) edge (a);
\draw [Arrow] (in) -- (a);
\draw [Arrow] (a) -- (out);
\end{scope}

\begin{scope}[shift={(1.9236,-1.7856)}, scale=0.5, transform shape, rotate=0]  
\node [inode] (a) at (-1.025,2.75) {};
\node [inode] (b) at (-1,1.3588) {};
\node [hidden] (in) at (-1.525,2.125) {};
\node [hidden] (out) at (-0.4712,2.1688) {};

\draw [edge,] (a) edge (b);
\draw [Arrow] (in) -- (b);
\draw [Arrow] (b) -- (out);
\end{scope}

\begin{scope}[shift={(4.9722,-1.81)}, scale=0.5, transform shape, rotate=0]  
\node [inode] (a) at (-1.025,2.75) {};
\node [inode] (b) at (-1,1.3588) {};
\node [hidden] (in) at (-1.525,3.525) {};
\node [hidden] (out) at (-0.53,2.21) {};

\draw [edge] (a) edge (b);
\draw [Arrow] (in) -- (a);
\draw [Arrow] (b) -- (out);
\end{scope}

\begin{scope}[shift={(5.9824,-1.8032)}, scale=0.5, transform shape]  
\node [inode] (a) at (-1.025,2.75) {};
\node [inode] (b) at (-1,1.3588) {};
\node [hidden] (in) at (-1.525,3.525) {};
\node [hidden] (out) at (-0.4712,2.1688) {};

\draw [edge, <->] (b) edge (a);
\draw [Arrow] (in) -- (a);
\draw [Arrow] (b) -- (out);
\end{scope}
\node [textnode] at (3.05,-1.55) {\footnotesize\sffamily{Rule}};

\end{tikzpicture}
    \begin{tikzpicture}
    \begin{customlegend}[ 
    legend columns=2,
    legend style={
    draw=none,
    column sep=2ex,
    font=\footnotesize\sffamily,
  },
  legend entries={Frequency, Edit Cost per Node},
  ]
    \addlegendimage{thick, blue, fill=blue!30, mark=square*}
    \addlegendimage{thick, red, fill=red!30,mark=*}

    \end{customlegend}
\end{tikzpicture}
    \caption{Most frequently extracted rules from real-world graphs. The red line indicates the total edit cost (number of edges added or deleted) per node over the course of the extractions with that rule.}
    \label{fig:real_world_rules}
    \vspace{-.5cm}
\end{figure}
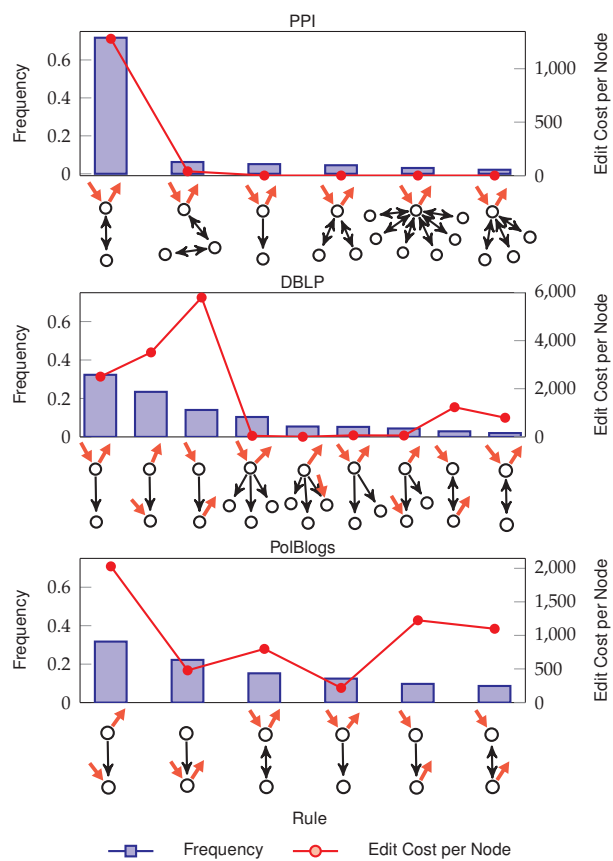

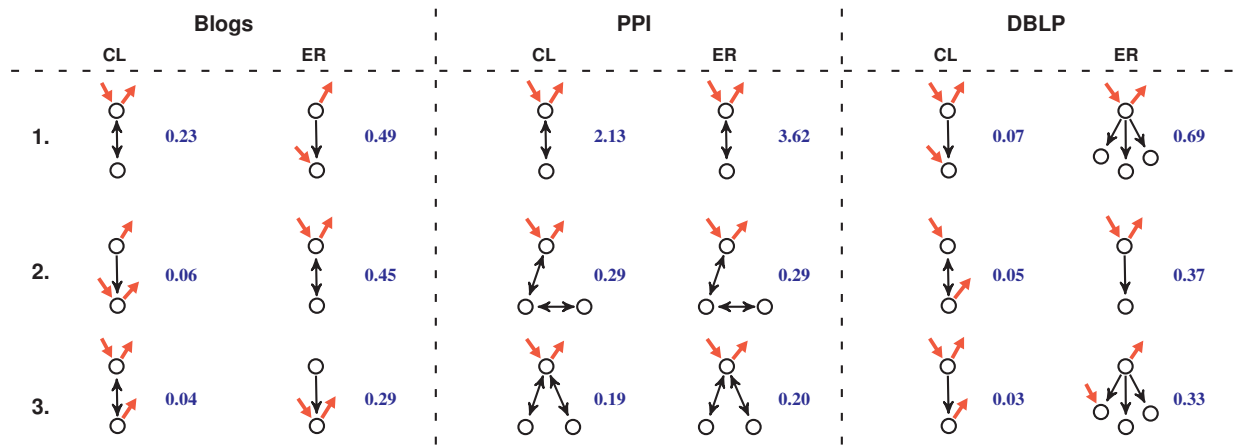
\begin{figure*}
    \centering
    \begin{tikzpicture}[scale=0.6, transform shape]

\begin{scope}[shift={(9.4,15)}, scale=0.95, transform shape]  

\node [inode] (a) at (-1.025,2.75) {};
\node [inode] (b) at (-1,1.3588) {};
\node [hidden] (in) at (-1.4373,3.525) {};
\node [hidden] (out) at (-0.4835,3.4811) {};

\draw [edge, <->] (a) edge (b);
\draw [Arrow] (in) -- (a);
\draw [Arrow] (a) -- (out);
\node [blue] (up) at (0.4861,2.1754) {\textbf{\Huge\large 0.23}};
\end{scope}

\begin{scope}[shift={(13.8165,15)}, scale=0.95, transform shape]  
\node [inode] (a) at (-1.025,2.75) {};
\node [inode] (b) at (-1,1.3588) {};
\node [hidden] (in) at (-1.6127,2.0341) {};
\node [hidden] (out) at (-0.5712,3.5688) {};

\draw [edge] (a) edge (b);
\draw [Arrow] (in) -- (b);
\draw [Arrow] (a) -- (out);

\node [blue] (up) at (0.4861,2.1754) {\textbf{\Huge\large 0.49}};
\end{scope}

\begin{scope}[shift={(18.8998,15)}, scale=0.95, transform shape]  
\node [inode] (a) at (-1.025,2.75) {};
\node [inode] (b) at (-1,1.3384) {};
\node [hidden] (in) at (-1.5,3.5725) {};
\node [hidden] (out) at (-0.5251,3.5534) {};

\draw [edge, <->] (a) edge (b);
\draw [Arrow] (in) -- (a);
\draw [Arrow] (a) -- (out);

\node [blue] (up) at (0.4861,2.1754) {\textbf{\Huge\large 2.13}};
\end{scope}

\begin{scope}[shift={(22.8998,15)}, scale=0.95, transform shape]  
\node [inode] (a) at (-1.025,2.75) {};
\node [inode] (b) at (-1,1.3384) {};
\node [hidden] (in) at (-1.5,3.5725) {};
\node [hidden] (out) at (-0.5251,3.5534) {};

\draw [edge, <->] (a) edge (b);
\draw [Arrow] (in) -- (a);
\draw [Arrow] (a) -- (out);

\node [blue] (up) at (0.5738,2.1754) {\textbf{\Huge\large 3.62}};
\end{scope}

\begin{scope}[shift={(27.8165,15)}, scale=0.95, transform shape]  
\node [inode] (a) at (-1.025,2.75) {};
\node [inode] (b) at (-1,1.3588) {};
\node [hidden] (in) at (-1.525,3.525) {};
\node [hidden] (out2) at (-1.6127,2.0512) {};
\node [hidden] (out) at (-0.5712,3.5688) {};

\draw [edge] (a) edge (b);
\draw [Arrow] (in) -- (a);
\draw [Arrow] (a) -- (out);
\draw [Arrow] (out2) -- (b);
\node [blue] (up) at (0.3984,2.1754) {\textbf{\Huge\large 0.07}};
\end{scope}

\begin{scope}[shift={(31.7332,15)}, scale=0.95, transform shape]  
\node [inode] (a) at (-1,2.75) {};
\node [inode] (b) at (-0.9915,1.3348) {};
\node [hidden] (in) at (-1.5499,3.5138) {};
\node [hidden] (out) at (-0.4091,3.4794) {};
\node [inode] (c) at (-0.415,1.6496) {};
\node [inode] (d) at (-1.573,1.6847) {};

\draw [Arrow] (in) -- (a);
\draw [Arrow] (a) -- (out);
\draw [edge] (a) edge (b);
\draw [edge] (a) edge (c);
\draw [edge] (a) edge (d);

\node [blue] (up) at (0.4861,2.1754) {\textbf{\Huge\large 0.69}};
\end{scope}

\begin{scope}[shift={(9.4,12)}, scale=0.95, transform shape]  
\node [inode] (a) at (-1.025,2.75) {};
\node [inode] (b) at (-1,1.3588) {};
\node [hidden] (in) at (-1.525,2.1218) {};
\node [hidden] (out2) at (-0.3849,2.0512) {};
\node [hidden] (out) at (-0.5712,3.4811) {};

\draw [edge] (a) edge (b);
\draw [Arrow] (in) -- (b);
\draw [Arrow] (a) -- (out);
\draw [Arrow] (b) -- (out2);

\node [blue] (up) at (0.4861,2.0877) {\textbf{\Huge\large 0.06}};
\end{scope}

\begin{scope}[shift={(13.8165,12)}, scale=0.95, transform shape]  
\node [inode] (a) at (-1.025,2.75) {};
\node [inode] (b) at (-1,1.3588) {};
\node [hidden] (in) at (-1.525,3.525) {};
\node [hidden] (out) at (-0.5712,3.5688) {};

\draw [edge, <->] (b) edge (a);
\draw [Arrow] (in) -- (a);
\draw [Arrow] (a) -- (out);

\node [blue] (up) at (0.4861,2.0877) {\textbf{\Huge\large 0.45}};
\end{scope}

\begin{scope}[shift={(18.8998,12)}, scale=0.95, transform shape]  
\node [inode] (a) at (-1,2.75) {};
\node [inode] (b) at (-0.1145,1.3348) {};
\node [hidden] (in) at (-1.5499,3.5138) {};
\node [hidden] (out) at (-0.4968,3.4794) {};

\node [inode] (d) at (-1.4853,1.3339) {};

\draw [Arrow] (in) -- (a);
\draw [Arrow] (a) -- (out);
\draw [edge, <->] (a) edge (d);
\draw [edge, <->] (b) edge (d);

\node [blue] (up) at (0.4861,2.0877) {\textbf{\Huge\large 0.29}};
\end{scope}

\begin{scope}[shift={(22.8998,12)}, scale=0.95, transform shape]  
\node [inode] (a) at (-1,2.75) {};
\node [inode] (b) at (-0.1145,1.3348) {};
\node [hidden] (in) at (-1.5499,3.5138) {};
\node [hidden] (out) at (-0.4091,3.4794) {};

\node [inode] (d) at (-1.4853,1.3339) {};

\draw [Arrow] (in) -- (a);
\draw [Arrow] (a) -- (out);
\draw [edge, <->] (a) edge (d);
\draw [edge, <->] (b) edge (d);

\node [blue] (up) at (0.5738,2.0877) {\textbf{\Huge\large 0.29}};
\end{scope}

\begin{scope}[shift={(27.8165,12)}, scale=0.95, transform shape]  
\node [inode] (a) at (-1.025,2.75) {};
\node [inode] (b) at (-1,1.3588) {};
\node [hidden] (in) at (-1.525,3.525) {};
\node [hidden] (out2) at (-0.3849,2.1389) {};

\draw [edge, <->] (b) edge (a);
\draw [Arrow] (in) -- (a);
\draw [Arrow] (b) -- (out2);

\node [blue] (up) at (0.3984,2.0877) {\textbf{\Huge\large 0.05}};
\end{scope}

\begin{scope}[shift={(31.7332,12)}, scale=0.95, transform shape]  
\node [inode] (a) at (-1.025,2.75) {};
\node [inode] (b) at (-1,1.3384) {};
\node [hidden] (in) at (-1.5,3.5725) {};
\node [hidden] (out) at (-0.5251,3.5534) {};

\draw [edge] (a) edge (b);
\draw [Arrow] (in) -- (a);
\draw [Arrow] (a) -- (out);

\node [blue] (up) at (0.4861,2.0877) {\textbf{\Huge\large 0.37}};
\end{scope}

\begin{scope}[shift={(9.4,9.3332)}, scale=0.95, transform shape]  
\node [inode] (a) at (-1.025,2.75) {};
\node [inode] (b) at (-1,1.3588) {};
\node [hidden] (in) at (-1.4373,3.525) {};
\node [hidden] (out2) at (-0.4726,2.1389) {};
\node [hidden] (out) at (-0.5712,3.4811) {};

\draw [edge, <->] (b) edge (a);
\draw [Arrow] (in) -- (a);
\draw [Arrow] (a) -- (out);
\draw [Arrow] (b) -- (out2);

\node [blue] (up) at (0.4861,2) {\textbf{\Huge\large 0.04}};
\end{scope}

\begin{scope}[shift={(13.8165,9.3332)}, scale=0.95, transform shape]  
\node [inode] (a) at (-1.025,2.75) {};
\node [inode] (b) at (-1,1.3588) {};
\node [hidden] (in) at (-1.525,2.1218) {};
\node [hidden] (out2) at (-0.4835,2.1827) {};

\draw [edge] (a) edge (b);
\draw [Arrow] (in) -- (b);
\draw [Arrow] (b) -- (out2);

\node [blue] (up) at (0.4861,2) {\textbf{\Huge\large 0.29}};
\end{scope}

\begin{scope}[shift={(18.8998,9.3332)}, scale=0.95, transform shape]  
\node [inode] (a) at (-1,2.75) {};
\node [inode] (b) at (-0.3776,1.3348) {};
\node [hidden] (in) at (-1.5499,3.5138) {};
\node [hidden] (out) at (-0.4968,3.4794) {};
\node [inode] (d) at (-1.4853,1.3339) {};

\draw [Arrow] (in) -- (a);
\draw [Arrow] (a) -- (out);
\draw [edge, <->] (a) edge (b);
\draw [edge, <->] (a) edge (d);

\node [blue] (up) at (0.4861,2) {\textbf{\Huge\large 0.19}};
\end{scope}

\begin{scope}[shift={(22.8998,9.3332)}, scale=0.95, transform shape]  
\node [inode] (a) at (-1,2.75) {};
\node [inode] (b) at (-0.3776,1.3348) {};
\node [hidden] (in) at (-1.5499,3.5138) {};
\node [hidden] (out) at (-0.4091,3.4794) {};
\node [inode] (d) at (-1.4853,1.3339) {};

\draw [Arrow] (in) -- (a);
\draw [Arrow] (a) -- (out);
\draw [edge, <->] (a) edge (b);
\draw [edge, <->] (a) edge (d);

\node [blue] (up) at (0.5738,2) {\textbf{\Huge\large 0.20}};
\end{scope}

\begin{scope}[shift={(27.8165,9.3332)}, scale=0.95, transform shape]  
\node [inode] (a) at (-1.025,2.75) {};
\node [inode] (b) at (-1,1.3588) {};
\node [hidden] (in) at (-1.525,3.525) {};
\node [hidden] (out2) at (-0.4726,2.1389) {};
\node [hidden] (out) at (-0.5712,3.5688) {};

\draw [edge,] (a) edge (b);
\draw [Arrow] (in) -- (a);
\draw [Arrow] (a) -- (out);
\draw [Arrow] (b) -- (out2);

\node [blue] (up) at (0.3984,2) {\textbf{\Huge\large 0.03}};
\end{scope}

\begin{scope}[shift={(31.7332,9.3332)}, scale=0.95, transform shape]  
\node [inode] (a) at (-1,2.75) {};
\node [inode] (b) at (-0.9915,1.3348) {};
\node [hidden] (in) at (-1.9884,2.4614) {};
\node [hidden] (out) at (-0.4968,3.4794) {};
\node [inode] (c) at (-0.415,1.6496) {};
\node [inode] (d) at (-1.573,1.6847) {};

\draw [Arrow] (in) -- (d);
\draw [Arrow] (a) -- (out);
\draw [edge] (a) edge (b);
\draw [edge] (a) edge (c);
\draw [edge] (a) edge (d);

\node [blue] (up) at (0.4861,2) {\textbf{\Huge\large 0.33}};
\end{scope}


\node [textnode,align=center, text width=2cm] at (6.75,17.0) {\LARGE{\textbf{\textsf{1.}}}};
\node [textnode,align=center, text width=2cm] at (6.75,14.0) {\LARGE{\textbf{\textsf{2.}}}};
\node [textnode,align=center, text width=2cm] at (6.75,11.0) {\LARGE{\textbf{\textsf{3.}}}};

\node [textnode] at (10.8,19.5) {\LARGE{\textbf{\textsf{Blogs}}}};
\node [textnode] at (19.9,19.5) {\LARGE{\textbf{\textsf{PPI}}}};
\node [textnode] at (28.8,19.5) {\LARGE{\textbf{\textsf{DBLP}}}};

\node [textnode] at (26.8,18.8) {\Large{\textbf{\textsf{CL}}}};
\node [textnode] at (30.8,18.8) {\Large{\textbf{\textsf{ER}}}};

\node [textnode] at (17.9,18.8) {\Large{\textbf{\textsf{CL}}}};
\node [textnode] at (21.9,18.8) {\Large{\textbf{\textsf{ER}}}};

\node [textnode] at (8.3835,18.8) {\Large{\textbf{\textsf{CL}}}};
\node [textnode] at (12.8,18.8) {\Large{\textbf{\textsf{ER}}}};

\node [hidden] (v42) at (5.9673,18.4976) {};
\node [hidden] (v43) at (33.5847,18.4976) {};
\draw[thick, loosely dashed]  (v42) edge (v43);

\node [hidden] (v1) at (15.5,19.9996) {};
\node [hidden] (v2) at (15.5,10.05) {};
\draw[thick, loosely dashed]  (v1) edge (v2);

\node [hidden] (v1) at (24.4956,19.9996) {};
\node [hidden] (v2) at (24.4956,10.05) {};
\draw[thick, loosely dashed]  (v1) edge (v2);

\end{tikzpicture}
    \caption{Comparison of top 3 most ``interesting'' results when compared to Chung-Lu (CL) and Erdos-Renyi (ER) null models. BUGGE extracts KT-grammars that highlight certain dynamics of each dataset. Some patterns are well known, as in the bidirectional edges of PPI networks, others require careful inspection and further study by domain experts.} 
    \label{fig:results}
    \vspace{-.5cm}
\end{figure*}

\vspace{.2cm}
\noindent\textbf{Maayan Stelzl Protein-Protein Interaction (PPI) Graph.} For the PPI network, almost all of the rules BUGGE finds have bidirected edges, suggesting that if protein A interacts with protein B, then the reverse is true. This is indeed the case; 95\% of the connections in the original graph are bidirected.

The most frequently extracted rules are visualized in Fig.~\ref{fig:real_world_rules}(top). By far the most frequent is a two-node rule where one node has boundary edges, and the other does not. However, we find that in the course of 866 total extractions, 2561 edges were deleted (denoted by the red line). Thus, the node lacking edges in the rule typically had a few edges \text{which were not held in common with its neighbor} in the real graph. This suggests that the general structure of the graph is to have proteins with very few interactions (spokes) connect to proteins with very many (hubs). We especially see this ``hub'' trait in some of the other top rules illustrated in Fig.~\ref{fig:real_world_rules}.


\vspace{.2cm}
\noindent\textbf{DBLP Article Citation Network.} For the DBLP citation network, BUGGE extracts 9 rules which are used the most frequently. They are illustrated in Fig.~\ref{fig:real_world_rules}(middle); many of which are similar to each other. As expected for a citation network, which should be a DAG, the most popular rules do not have bidirected edges.


We observed that in all of these rules, at most one node has outgoing boundary edges and at most one node has incoming. This means that for most pairs of connected nodes, it was cheapest for BUGGE to \textit{delete} all but one node's in edges and \textit{delete} all but one node's out edges. This, in turn, means that for most pairs of connected nodes, they had more distinct edges than edges in common. In terms of citations, this means a pair of articles connected by a citation are more likely to cite and be cited by different articles than by the same ones. This level of expressibility is exactly what we seek; we, therefore, encourage domain experts in library sciences (or proteomics or the social Web) to investigate these findings further. 



\vspace{.2cm}
\noindent\textbf{Moreno Blogs-Blogs Network.} The Blogs network is another form of a citation network, but because multiple articles on the same blog count as the same node and two blogs can frequently cite one another, the Blogs network will not be nearly as DAG-like. Cycles and mutual citations should be much more common. We expected the blogs-to-blogs graph to have much less regular structure due to their complex social dynamics. However, we did obtain some of the same observations as in the DBLP citation network. In particular, that the shared citations between two blogs are fewer than the distinct citations. 


\subsection*{Finding Interesting Rules}
These rule probabilities give a good indication of the structure of the graph. However, it could be that some rules are just more likely than others, especially within graphs of the same degree distribution. So, it is important that we find the rules that are most \textit{interesting} - not just most frequent. Defining what is ``interesting'' can be difficult; fortunately, null graph models are well suited for precisely this task.

For each real-world graph we create two null graph models: (1) an Erdos Renyi Random graph (ER) containing the same number of nodes and edges as the original graph, and (2) a random graph that matches the original graph's degree distribution using a directed version of Chung-Lu's Configuration model (CL)~\cite{aiello2000random, newman2001random}. We use BUGGE to extract a KT-grammar from the two null models for each real-world graph.

The extracted KT-grammars are a distribution of rules. So we can compare the graph models using KL-Divergence to determine how similar they are:

\begin{equation*}
    KL(p,q) = -\sum_{R\in \{G\cup G^{\emptyset}\}} p(R) \log \frac{q(R)}{p(R)}
\end{equation*}

\noindent where $G$ is the KT-grammar extracted from the original graph, and $G^{\emptyset}$ is the KT-grammar extracted from the null model, either ER or CL; $p(R)$ and $q(R)$ are the probabilities that $R$ appears in the grammar extracted from the original graph and the null model respectively. In some cases a rule may not appear in both graphs, so we perform Laplacian smoothing on these distributions to avoid errors caused by dividing by zero.

The KL divergence result itself is not particularly meaningful, however, the contribution of each rule $R$ to the overall result represents the relative difference in their occurrence. Therefore, we rank each rule's contribution to the overall KL divergence and illustrate the top 3 rules in Fig.~\ref{fig:results} for comparisons of real-world datasets against the null models.

Many aspects of our results could be commented on. We will highlight a few: The frequency of rules with bidirected edges in Figure~\ref{fig:results} shows that neither the degree distribution nor the ER model capture these relationships. In the Blogs vs. CL comparison, we see that even though (as discussed earlier) most of the extracted rules do not have multiple out edges, they are more common in the original graph than the degree distribution alone would dictate. In the DBLP citation graph vs. ER, we see that BUGGE finds the original graph has much more tree-like/DAG-like rules. 

\nop{
\begin{figure}[t]
    \centering
    \begin{tikzpicture}
\begin{groupplot}[
    group style={
        group size=1 by 1,
        xlabels at=edge bottom,
        xticklabels at=edge bottom,
        vertical sep=5pt,
        horizontal sep=10pt,
    },
    width  = 3in,
    height = 1.55in,
    xlabel = {},
    axis line style={-},
    legend style={
    	cells={align=right},
    	legend columns=4,
    	column sep=6pt,
    	font=\tiny
    },
]
\nop{
\nextgroupplot[
    ylabel={\footnotesize\sffamily{Time (log s)}},
    ymin=1, 
    symbolic x coords={bin tree,tree ring,ring lat,protein,blogs,dblp}, 
    xticklabels={,,},
    enlargelimits=true,
    ymajorgrids = true,
    xtick=data,
    ybar,
    ymin=1,
    /pgf/bar width=3pt,
    ymode=log,
    xtick style={draw=none}
    ]
    
    \addplot[thick, draw=red, fill=red!30] coordinates  
    {
        (bin tree,0.57)
        (tree ring,0.96)
        (ring lat,8.11)
        (protein,0.63)
        (blogs,21.53)
        (dblp,309.39)
    };
     \addlegendentry{SUBDUE};
    
    \addplot[thick, draw=blue, fill=blue!30] coordinates 
    {
        (bin tree,4.2)
        (tree ring,6.4211)
        (ring lat,6.461)
        (protein,1.6)
        (blogs,2.4742)
        (dblp,116.45)
    };
    
    \addplot[thick, draw=green!50!black, fill=green!30] coordinates 
    {
        (bin tree,17.2)
        (tree ring,16.3)
        (ring lat,18.1)
        (protein,14.9)
        (blogs,18.1)
        (dblp,333.4)
    };
    
    \addplot[thick, draw=black, fill=gray!50!white] coordinates 
    {
        (bin tree,284.472)
        (tree ring,516.323)
        (ring lat,1071.641)
        (protein,1193.6)
        (blogs,5168.8)
        (dblp,64687.3)
    };
    
    }

\nextgroupplot[
    ylabel={\footnotesize\sffamily{Compression Rate}},
    ymin=0, ymax=1,
    symbolic x coords={bin tree,tree ring,ring lat,protein,blogs,dblp}, 
    xticklabels={Binary Tree, Tree of Rings,Ring Lattice,PPI,Blogs,DBLP},
    xticklabel style={
        rotate=90,
        font=\footnotesize\sffamily,
    },
    enlargelimits=true,
    ymajorgrids = true,
    xtick=data,
    ybar,
    /pgf/bar width=3pt,
    xtick style={draw=none}
    ]
    
    \addplot[thick, draw=red, fill=red!30] coordinates  
    {
        (bin tree,0.51)
        (tree ring,0.43)
        (ring lat,0.50)
        (protein,-0.07)
        (blogs,0.03)
        (dblp,0.0)
    };
    
    \addplot[thick, draw=blue, fill=blue!30] coordinates 
    {
        (bin tree,0.99)
        (tree ring,1.0)
        (ring lat,0.99)
        (protein,0.36)
        (blogs,0.57)
        (dblp,0.30)
    };
    
    \addplot[thick, draw=green!50!black, fill=green!30] coordinates 
    {
        (bin tree,0.0)
        (tree ring,0.0)
        (ring lat,0.0)
        (protein,0.08)
        (blogs,0.12)
        (dblp,0.07)
    };
    
    \addplot[thick, draw=black, fill=gray!50!white] coordinates 
    {
        (bin tree,0.84)
        (tree ring,0.82)
        (ring lat,0.81)
        (protein,0.28)
        (blogs,0.01)
        (dblp,0.03)
    };
    
\end{groupplot}
\end{tikzpicture}
    \begin{tikzpicture}
    \begin{customlegend}[ 
    legend columns=4,
    legend style={
    draw=none,
    column sep=2ex,
    font=\footnotesize\sffamily,
  },
  legend entries={SUBDUE, CNRG, VoG, BUGGE},
  ]
    \addlegendimage{thick, red, fill=red!30, mark=square*}
    \addlegendimage{thick, blue, fill=blue!30, mark=square*}
    \addlegendimage{thick, green, fill=green!30, mark=square*}
    \addlegendimage{thick, black, fill=gray!50, mark=square*}
    \end{customlegend}
\end{tikzpicture}
    \caption{Compression achieved on synthetic and real graphs (higher is better). Higher compression indicates a greater likelihood that the system found structures relevant to the graph. Note that both SUBDUE and CNRG provide lossy compressions, so their numbers are not \textit{directly} comparable to VoG or BUGGE. BUGGE, like all systems except for VoG, does better on the synthetic graphs since there is more structure to discover. }
    \label{fig:runtime}
\end{figure}
}

\subsection*{Performance Comparisons} Finally, a direct comparison of the compression rates of BUGGE, SUBDUE, CNRG, and VoG is problematic. CNRG and SUBDUE are lossy models, while BUGGE and VoG are lossless models. 

Likewise, direct runtime comparisons are also problematic. For example, the default settings for SUBDUE search for grammars of arbitrary size, which does not scale to even medium-sized graphs; so we its max structure size to 8. Each algorithm is written in different programming languages using different graph libraries, etc. Runtimes in comparisons ranged from less than a minute on the smallest graphs to around 18 hours for BUGGE on the largest real-world graph.

\section{Conclusions}

The present work describes BUGGE: the Bottom-Up Graph Grammar Extractor, which extracts grammar rules that represent interpretable substructures from large graph data sets. Using synthetic data sets we explored the expressivity of these grammars and showed that they clearly articulated the specific dynamics that generated the synthetic data. On real-world data sets, we further explored the more frequent and most interesting (from an information-theoretic point of view) rules and found that they clearly represent meaningful substructures that may be useful to domain experts. This level of expressivity and interpretability is needed in many fields with large and complex graph data. So, we repeat our call for domain experts to investigate these findings further.

In future work, we intend to focus on extending these formalisms to cover temporal/evolving graphs, like the work done in synchronous HRGs~\cite{pennycuff2018synchronous} and temporal motifs~\cite{paranjape2017motifs}. It is also likely that the KT-grammars extracted here can be used to generate faithful null models of a graph.

\vspace{.2cm}
\noindent\textbf{Acknowledgements.} This research is supported by a grant from the US National Science Foundation (\#1652492). 

\bibliographystyle{./bibliography/IEEEtran}
\bibliography{./bibliography/references}

\begin{thebibliography}{10}
\providecommand{\url}[1]{#1}
\csname url@samestyle\endcsname
\providecommand{\newblock}{\relax}
\providecommand{\bibinfo}[2]{#2}
\providecommand{\BIBentrySTDinterwordspacing}{\spaceskip=0pt\relax}
\providecommand{\BIBentryALTinterwordstretchfactor}{4}
\providecommand{\BIBentryALTinterwordspacing}{\spaceskip=\fontdimen2\font plus
\BIBentryALTinterwordstretchfactor\fontdimen3\font minus
  \fontdimen4\font\relax}
\providecommand{\BIBforeignlanguage}[2]{{%
\expandafter\ifx\csname l@#1\endcsname\relax
\typeout{** WARNING: IEEEtran.bst: No hyphenation pattern has been}%
\typeout{** loaded for the language `#1'. Using the pattern for}%
\typeout{** the default language instead.}%
\else
\language=\csname l@#1\endcsname
\fi
#2}}
\providecommand{\BIBdecl}{\relax}
\BIBdecl

\bibitem{ahmed2015efficient}
N.~K. Ahmed, J.~Neville, R.~A. Rossi, and N.~Duffield, ``Efficient graphlet
  counting for large networks,'' in \emph{ICDM}.\hskip 1em plus 0.5em minus
  0.4em\relax IEEE, 2015, pp. 1--10.

\bibitem{seshadhri2012community}
C.~Seshadhri, T.~G. Kolda, and A.~Pinar, ``Community structure and scale-free
  collections of erd{\H{o}}s-r{\'e}nyi graphs,'' \emph{Physical Review E},
  vol.~85, no.~5, p. 056109, 2012.

\bibitem{aguinaga2016growing}
S.~Agui{\~n}aga, R.~Palacios, D.~Chiang, and T.~Weninger, ``Growing graphs from
  hyperedge replacement graph grammars,'' in \emph{CIKM}.\hskip 1em plus 0.5em
  minus 0.4em\relax ACM, 2016, pp. 469--478.

\bibitem{aguinaga2018learning}
S.~Aguinaga, D.~Chiang, and T.~Weninger, ``Learning hyperedge replacement
  grammars for graph generation,'' \emph{IEEE Trans. on Pattern Analysis and
  Machine Intelligence}, vol.~41, pp. 625--638, 2019.

\bibitem{koutra2014vog}
D.~Koutra, U.~Kang, J.~Vreeken, and C.~Faloutsos, ``Vog: Summarizing and
  understanding large graphs,'' in \emph{SDM}.\hskip 1em plus 0.5em minus
  0.4em\relax SIAM, 2014, pp. 91--99.

\bibitem{yan2002gspan}
X.~Yan and J.~Han, ``gspan: Graph-based substructure pattern mining,'' in
  \emph{ICDM}.\hskip 1em plus 0.5em minus 0.4em\relax IEEE, 2002, pp. 721--724.

\bibitem{yan2003closegraph}
------, ``Closegraph: mining closed frequent graph patterns,'' in
  \emph{Proceedings of the ninth ACM SIGKDD international conference on
  Knowledge Discovery and Data Mining}.\hskip 1em plus 0.5em minus 0.4em\relax
  ACM, 2003, pp. 286--295.

\bibitem{holder1994substucture}
L.~B. Holder, D.~J. Cook, S.~Djoko \emph{et~al.}, ``Substucture discovery in
  the subdue system.'' in \emph{SIGKDD}, 1994, pp. 169--180.

\bibitem{gudkov2016generalized}
V.~Gudkov, ``Generalized entropies of complex and random networks,''
  \emph{Mathematical Foundations and Applications of Graph Entropy}, vol.~6,
  pp. 41--61, 2016.

\bibitem{sikdar2019modeling}
S.~Sikdar, J.~Hibshman, and T.~Weninger, ``Modeling graphs with vertex
  replacement grammars,'' in \emph{ICDM}.\hskip 1em plus 0.5em minus
  0.4em\relax IEEE, 2019.

\bibitem{reddy2019edge}
R.~Reddy, S.~Chandar, and B.~Ravindran, ``Edge replacement grammars: A formal
  language approach for generating graphs,'' in \emph{SDM}.\hskip 1em plus
  0.5em minus 0.4em\relax SIAM, 2019, pp. 351--359.

\bibitem{kemp2008discovery}
C.~Kemp and J.~B. Tenenbaum, ``The discovery of structural form,'' \emph{PNAS},
  vol. 105, no.~31, pp. 10\,687--10\,692, 2008.

\bibitem{ehrig1999handbook}
H.~Ehrig, G.~Rozenberg, and H.-J. rg~Kreowski, \emph{Handbook of graph grammars
  and computing by graph transformation}.\hskip 1em plus 0.5em minus
  0.4em\relax World Scientific, 1999, vol.~3.

\bibitem{elias1975universal}
P.~Elias, ``Universal codeword sets and representations of the integers,''
  \emph{IEEE Trans. on Information Theory}, vol.~21, no.~2, pp. 194--203, 1975.

\bibitem{avis1996reverse}
D.~Avis and K.~Fukuda, ``Reverse search for enumeration,'' \emph{Discrete
  Applied Mathematics}, vol.~65, no. 1-3, pp. 21--46, 1996.

\bibitem{robins2007introduction}
G.~Robins, P.~Pattison, Y.~Kalish, and D.~Lusher, ``An introduction to
  exponential random graph (p*) models for social networks,'' \emph{Social
  Networks}, vol.~29, no.~2, pp. 173--191, 2007.

\bibitem{simonovsky2018graphvae}
M.~Simonovsky and N.~Komodakis, ``Graphvae: Towards generation of small graphs
  using variational autoencoders,'' in \emph{International Conference on
  Artificial Neural Networks}.\hskip 1em plus 0.5em minus 0.4em\relax Springer,
  2018, pp. 412--422.

\bibitem{you2018graphrnn}
J.~You, R.~Ying, X.~Ren, W.~L. Hamilton, and J.~Leskovec, ``Graphrnn:
  Generating realistic graphs with deep auto-regressive models,'' \emph{arXiv
  preprint arXiv:1802.08773}, 2018.

\bibitem{bojchevski2018netgan}
A.~Bojchevski, O.~Shchur, D.~Z{\"u}gner, and S.~G{\"u}nnemann, ``Netgan:
  Generating graphs via random walks,'' \emph{arXiv preprint arXiv:1803.00816},
  2018.

\bibitem{tang2015line}
J.~Tang, M.~Qu, M.~Wang, M.~Zhang, J.~Yan, and Q.~Mei, ``Line: Large-scale
  information network embedding,'' in \emph{WWW}, 2015, pp. 1067--1077.

\bibitem{grover2016node2vec}
A.~Grover and J.~Leskovec, ``node2vec: Scalable feature learning for
  networks,'' in \emph{SIGKDD}.\hskip 1em plus 0.5em minus 0.4em\relax ACM,
  2016, pp. 855--864.

\bibitem{kipf2016variational}
T.~N. Kipf and M.~Welling, ``Variational graph auto-encoders,'' \emph{arXiv
  preprint arXiv:1611.07308}, 2016.

\bibitem{goyal2018graph}
P.~Goyal and E.~Ferrara, ``Graph embedding techniques, applications, and
  performance: A survey,'' \emph{Knowledge-Based Systems}, vol. 151, pp.
  78--94, 2018.

\bibitem{sarajlic2016graphlet}
A.~Sarajli{\'c}, N.~Malod-Dognin, {\"O}.~N. Yavero{\u{g}}lu, and
  N.~Pr{\v{z}}ulj, ``Graphlet-based characterization of directed networks,''
  \emph{Scientific reports}, vol.~6, p. 35098, 2016.

\bibitem{milo2002network}
R.~Milo, S.~Shen-Orr, S.~Itzkovitz, N.~Kashtan, D.~Chklovskii, and U.~Alon,
  ``Network motifs: simple building blocks of complex networks,''
  \emph{Science}, vol. 298, no. 5594, pp. 824--827, 2002.

\bibitem{aiello2000random}
W.~Aiello, F.~Chung, and L.~Lu, ``A random graph model for massive graphs,'' in
  \emph{STOC}.\hskip 1em plus 0.5em minus 0.4em\relax Acm, 2000, pp. 171--180.

\bibitem{newman2001random}
M.~E. Newman, S.~H. Strogatz, and D.~J. Watts, ``Random graphs with arbitrary
  degree distributions and their applications,'' \emph{Physical review E},
  vol.~64, no.~2, p. 026118, 2001.

\bibitem{pennycuff2018synchronous}
C.~Pennycuff, S.~Sikdar, C.~Vajiac, D.~Chiang, and T.~Weninger, ``Synchronous
  hyperedge replacement graph grammars,'' in \emph{International Conference on
  Graph Transformation}.\hskip 1em plus 0.5em minus 0.4em\relax Springer, 2018,
  pp. 20--36.

\bibitem{paranjape2017motifs}
A.~Paranjape, A.~R. Benson, and J.~Leskovec, ``Motifs in temporal networks,''
  in \emph{WSDM}.\hskip 1em plus 0.5em minus 0.4em\relax ACM, 2017, pp.
  601--610.

\end{thebibliography}

\begin{appendices}


\appendix\label{appendix-a}

\section{\textbf{Appendix - Encoding Scheme}}

When compressing a graph, we use three encodings: A \textbf{graph encoding} (stores a minimalist adjacency list), a \textbf{grammar rule encoding}, and an \textbf{application encoding} (stores a sequence of applications of the grammar rules to a graph).

\vspace{.2cm}
\noindent\textbf{Graph Encoding.} Given a directed graph $H = (V, E)$, the number of bits $B_H$ it takes to encode $G$ is:
\begin{equation*}
    B_H = (2\lceil \log_2 |V| \rceil - 1) + |V| + |E|(\lceil \log_2 |V| \rceil + 1)
\end{equation*}

\vspace{.2cm}
\noindent\textbf{Grammar Encoding.} A grammar rule is basically a graph with additional boundary information. 
%
%
The total number of bits $B_{R_k}$ to encode a grammar rule with $k$ nodes is:

\begin{equation*}
    B_{R_k} = \lceil \log_2 |V| \rceil + k(\lceil \log_2 k \rceil + 2) + k(k-1) + 1
\end{equation*}

\vspace{.2cm}
\noindent\textbf{Application Encoding.} An application encoding consists of a sequence of instructions for applying grammar rules. These instructions include an id number of the rule to apply, the id of the node to apply the rule to, and information concerning any edges which were added or deleted during the extraction process.
%
%
%
The bits $B_{A_{km}}$ to record the application of a $k$-node rule with $m$ edge approximations takes:

\begin{align*}
    B_{A_{km}} = 2 + \lceil \log_2 |V| \rceil + m(\lceil \log_2 k \rceil + \lceil \log_2 |V| \rceil + 1) \nonumber \\
    + \begin{cases} 
      \lceil \log_2 |V| \rceil & \text{different rule used before} \\
      0 & \text{same rule used before}
   \end{cases}
\end{align*}



\end{appendices}

\end{document}